\documentclass{article}
\usepackage{graphicx}  % standard LaTeX graphics tool;
\usepackage{amsmath}
\usepackage{subcaption}% For getting proper math eqns
\usepackage{amssymb}   % Used for getting various symbols eg. gtrsim, lesssim etc.
\usepackage{bm} % For bold math (esp of lower greek letters)
\usepackage{dcolumn}% Align table columns on decimal point
\usepackage{color}
\usepackage{mathrsfs}
\usepackage{amsfonts}
\usepackage{varioref}
\usepackage[margin=1.6in]{geometry}
\usepackage{slashed}
\RequirePackage[colorlinks,citecolor=blue,urlcolor=magenta,linkcolor=blue]{hyperref}
\def\p{\partial}
\def\e{\epsilon}

\def\be{\begin{equation}}
\def\ee{\end{equation}}

\labelformat{section}{Section #1} 
\labelformat{subsection}{Section #1} 
\labelformat{subsubsection}{Section #1}
\labelformat{subsubsubsection}{Section #1}
\labelformat{equation}{Eq.~(#1)} 
\labelformat{figure}{Fig.~#1} 
\labelformat{subfigure}{Fig.~\thefigure#1} 
\labelformat{table}{Tab.~#1} 
\labelformat{appendix}{Appendix #1}
\title{\bf Loop correction and resummation of vertex functions for a  self interacting  scalar field  in the  de Sitter spacetime}
\author{$^{1}$Sourav Bhattacharya\footnote{sbhatta.physics@jadavpuruniversity.in}\,\, and $^2$Sudesh Kumar\footnote{sudesh.21phz0007@iitrpr.ac.in}\\
\small{$^1$Relativity and Cosmology Research Centre, Department of Physics, Jadavpur University, Kolkata 700 032, India}\\
\small{$^2$Department of Physics, Indian Institute of Technology Ropar, Rupnagar, Punjab 140 001, India}\\}
\begin{document}
\maketitle
%%%%%%%%%%%%%%%%%%%%%%%%%%%%%%%%%%%%%%%%%%%%%%%%%%%%%%%%%%%%%%%
\begin{abstract}
\noindent
We consider a massless and minimally coupled self interacting quantum scalar field theory in the inflationary de Sitter background of dimension four. The self interaction potential is taken to be either  quartic, $\lambda \phi^4/4!$, or quartic plus cubic, $\lambda \phi^4/4!+\beta \phi^3/3!$ ($\lambda \,{\ensuremath >}\,0$). We compute the  four  and three point vertex functions up to two loop. The purely local or partly local part of these  renormalised  loop corrected vertex functions grow unboundedly  after sufficient number of de Sitter $e$-foldings, due to the appearances of secular logarithms. We focus on the purely local part of the vertex functions and attempt a resummation of them in terms of the dynamically  generated mass of the scalar field at late times. Such local logarithms have sub-leading powers compared to the non-local leading ones which can be resummed via the stochastic formalism.  The variation of these vertex functions are investigated with respect to the tree level couplings numerically.  Since neither the secular effect, nor the dynamical  generation of field mass is possible in the Minkowski spacetime, the above phenomenon has no flat spacetime analogue. We have also compared our result with the ones that could be found via the recently proposed renormalisation group techniques. All these results suggest that at late times the value of the non-perturbative vertex function should be less than the tree level coupling.
\end{abstract}
\vskip .5cm

\newpage

\tableofcontents

%%%%%%%%%%%%%%%%%%%%%%%%
\section{Introduction}\label{S1}
%%%%%%%%%%%%%%%%%%%%%%%

The standard cosmological model predicts  that our universe started expanding with a big bang, while the rate of expansion always decreasing monotonically. As such, this model has been very successful in predicting the redshift of galaxies, the cosmic microwave background radiation, the abundances of light elements and so on.    However, several other important issues like the  observed spatial flatness of our universe at present, the observed high degree of isotropy at very large scales and the rarity/unobservability of relics like the magnetic monopoles cannot be explained by this standard big bang model. The  primordial cosmic inflation is a proposed phase of very rapid, near exponential accelerated expansion of our very early universe,  in order to provide a possible solution to these problems. Indeed,  the inflationary phase not only offers satisfactory answers to these aforementioned puzzles, but also provides a robust mechanism  to generate primordial cosmological density perturbations, as the seed to the cosmic web we observe in the sky today at large scales. We refer our reader to~\cite{Wein, Mukhanov:2005sc} and references therein for excellent pedagogical discussion on these issues. 

In general relativity, the accelerated expansion of the spacetime always necessitates  some exotic matter field with positive energy density but negative, isotropic pressure, known as the dark energy.  The simplest form the dark energy is just the cosmological constant,  $\Lambda$. A slowly rolling scalar field off a potential can also work effectively as dark energy. However, such description lacks any microscopic quantum description, which can be large in a time dependent background. Now, the current observed density of dark energy is naturally much small compared to that of the primordial inflationary one. It turns out that only about $10\%$ mismatch of the current dark energy density would have led to drastic changes in the formation of large scale structures.  Thus it is an interesting question to ask,  how did $\Lambda$/dark energy attained today's tiny value? This is known as the cosmic coincidence puzzle~\cite{Tsamis, Ringeval}. An open issue in this context is : can the backreaction of the inflationary quantum  fluctuations, in particular the gravitons, screen $\Lambda$, e.g.~\cite{Miao:2021gic} and references therein. We also refer our reader to  e.g.~\cite{Dadhich, Padmanabhan, Alberte, Appleby, Khan:2022bxs, Evnin:2018zeo} for different aspects/proposals to solution of this  problem. 

In particular, a massless quantum  field theory with zero rest mass but not conformally invariant  (e.g., gravitons and massless minimally coupled scalars) can generate de Sitter breaking large loop corrections growing monotonically with time, proportional to the logarithm of the scale factor, known as the {\it secular effect}. This indicates that after sufficient number of $e$-foldings the perturbation theory breaks down, necessitating some resummation~\cite{Floratos}.  

Free quantum  field quantisation  in the de Sitter spacetime can be seen in~\cite{Chernikov:1968zm, Bunch:1978yq, Linde:1982uu, Starobinsky:1982ee, Allen:1985ux, Allen, Karakaya:2017evp}. It is well known by now that  a massless and  minimally coupled scalar field cannot have a  de Sitter invariant vacuum state or two point function.   For such a scalar with self interaction, the loops then naturally show de Sitter breaking effect growing monotonically with time  mentioned above, e.g.~\cite{Onemli:2002hr, Brunier:2004sb, Kahya:2009sz, Boyanovsky:2012qs, Onemli:2015pma, Prokopec:2003tm, Miao:2006pn, Prokopec:2007ak, Liao:2018sci, Miao:2020zeh, Glavan:2019uni, Karakaya:2019vwg, Cabrer:2007xm, Prokopec:2003qd, Boran:2017fsx} and references therein.  Efforts to resum such non-perturbative infrared effect (appears as  the logarithm of the scale factor) by various methods can be seen in e.g.~\cite{Miao:2021gic, Moreau:2018ena, Moreau:2018lmz, Gautier:2015pca, Serreau:2013eoa, Serreau:2013koa, Serreau:2013psa, Ferreira:2017ogo, Weinberg, Burgess:2009bs, Burgess:2015ajz, Youssef:2013by, Baumgart:2019clc, Kitamoto:2018dek, Kamenshchik:2020yyn, Kamenshchik:2021tjh, Tsamis:2005hd, Bhattacharya:2022aqi, Bhattacharya:2022wjl, Bhattacharya:2023yhx, Bhattacharya:2023xvd, Glavan:2021adm, Litos:2023nvj, Glavan:2023lvw, Glavan:2023tet, Miao:2024nsz}. We also refer our reader to~\cite{Starobinsky:1986fx, Starobinsky:1994bd} (see also~\cite{Cho:2015pwa, Prokopec:2015owa, Garbrecht:2013coa, Vennin:2015hra, Cruces:2022imf, Finelli:2008zg, Markkanen:2019kpv, Markkanen:2020bfc, Enqvist:2017kzh} for recent developments) for a stochastic formulation for resumming correlation functions at late times for self interacting scalar field theory  in the inflationary background. Perhaps one of the most important prediction of this resummation is the emergence of a dynamically generated scalar mass at late times~\cite{Starobinsky:1994bd, davis, Beneke:2012kn}. This is certainly a manifestation of strong quantum fluctuations. The dynamical mass may leave interesting footprints into the cosmic microwave background.   \\

\noindent
In this work we wish to consider secular effect in the context of vertex loop corrections in self interacting scalar field theories with $V(\phi)=\lambda \phi^4/4!$ and $V(\phi)= \lambda \phi^4/4!+\beta\phi^3/3!$ ($\lambda\,{\ensuremath >} \,0$), up to two loop order. For a quartic self interaction, the one loop correction to the four point vertex can be seen in~\cite{Brunier:2004sb}.  The pure vertex part of these  amplitudes essentially correspond to the purely local part of the corresponding amputated diagrams. By {\it local}, we essentially  refer to the part of the diagram which shrinks to a single point, by the virtue of relevant $\delta$-functions. In flat spacetime such local parts are solely divegent and renormalisable, whereas for de Sitter, they also contain non-divergent parts containing powers of the secular logarithm, which  cannot be renormalised away. We emphasise that these logarithms are {\it not} the leading infrared logarithms which the stochastic formalism resums successfully~\cite{Starobinsky:1994bd}. In any given diagram, the power of the leading logarithms equals to the total number of internal lines, whereas the aforementioned local logarithms are of subleading powers. In this work we wish to concern ourselves with such purely local, subleading logarithms only.     After briefly reviewing the necessary ingredients in the next section, we compute the renormalised  one and two loop corrections to the quartic and cubic local vertex functions respectively in \ref{s3}, \ref{s4} and \ref{A}, demonstrating explicitly the secular effect. We next attempt a bit  pedestrian approach to resum these perturbative expressions in \ref{s5} at late times, by trying to draw a correspondence between the local vertex functions and the local self energy, and hence the dynamically generated rest mass of the field at late times.     We shall also briefly discuss other recently proposed  resummation methods related to the renormalisation group and compare their predictions with ours in \ref{s5}. The resummed vertex function turns out to be less than the tree level coupling constant. Finally we conclude in \ref{s6}.\\

\noindent
 We shall work with the mostly positive signature of the metric in $d = 4 - \epsilon$ ($\epsilon$ = $0^{+}$) dimensions and will set $c = 1 = \hbar$ throughout.  Also for the sake of brevity, we shall denote for powers of propagators and logarithms respectively as, $(i\Delta)^n\equiv i\Delta^n$ and $(\ln x )^n \equiv \ln^n x$.

%%%%%%%%%%%%%
\section{A brief review of the basic ingredients}\label{s2}
%%%%%%%%%%%%%

\noindent
We wish to briefly outline below the framework we shall be working in, referring our reader to~\cite{Brunier:2004sb} for detail. We begin with the metric of the cosmological de Sitter spacetime 
\begin{eqnarray}
ds^2 = -dt^2 + a^2(t)d\vec{x} \cdot d\vec{x} = a^2(\eta) \left(-d\eta^2 +  d\vec{x} \cdot d\vec{x} \right)
\label{y0}
\end{eqnarray}
where the scale factor $a(t)$ equals $e^{Ht}$ with $H^2=\Lambda/3$, $\Lambda$ being the positive cosmological constant. $t$ and $\eta$ are respectively called the cosmological and conformal time, with $\eta= -e^{-Ht}/H$, thus  $a(\eta)=  -1/H\eta$. The de Sitter metric has a symmetry of time translation followed by a coordinate scaling. Utilising this symmetry, we set the temporal range $0 \leq t < \infty$, so that $-H^{-1}\leq \eta <0^-$.

We take the bare Lagrangian density to be that of a scalar field $\psi'$, with quartic and cubic self interaction potential
\begin{eqnarray}
S=\int \sqrt{-g} d^d x\left[ -\frac12 g^{\mu\nu} (\nabla_{\mu} \psi')(\nabla_{\nu} \psi') -\frac12 m_0^2 \psi'^2 - \frac{\lambda_0}{4!} \psi'^4 - \frac{\beta_0}{3!} \psi'^3 - \tau_0 \psi'    \right]
\label{y1}
\end{eqnarray}
We define the field strength renormalisation, $\phi=\psi'/\sqrt{Z}$, so that  
\begin{eqnarray}
S= \int d^d x\left[ -\frac{Z}{2} \eta^{\mu\nu} a^{d-2} (\p_{\mu} \phi)(\p_{\nu} \phi) -\frac{1}{2} Z m_0^2 \phi^2 a^d - \frac{Z^2 \lambda_0}{4!} \phi^4 a^d - \frac{\beta_0 Z^{3/2}}{3!} \phi^3 a^d - \tau_0 \sqrt{Z} \phi \,a^d \right]
\label{y2}
\end{eqnarray}
We take the rest mass of the field to be vanishing, so that 
\begin{eqnarray}
&&Z= 1+\delta Z \qquad Z m_0^2 = 0+\delta m^2 \qquad Z^2 \lambda_0 = \lambda +\delta \lambda \qquad \beta_0 Z^{3/2}= \beta+\delta \beta   \qquad \tau_0 \sqrt{Z} =\alpha   
\label{y3}
\end{eqnarray}
where $\delta m^2$, $\delta \lambda$, $\delta \beta$ and $\alpha$ are counterterms.
Substituting now the above into \ref{y2}, we have
\begin{eqnarray}
&&S=  \int d^d x \left[-\frac{1}{2} \eta^{\mu\nu} a^{d-2} (\p_{\mu} \phi)(\p_{\nu} \phi)  - \frac{\lambda}{4!} \phi^4 a^d - \frac{\beta}{3!} \phi^3 a^d \right.\nonumber\\&&\left.-\frac{\delta Z}{2} \eta^{\mu\nu} a^{d-2} (\p_{\mu} \phi)(\p_{\nu} \phi)  - \frac12 \delta m^2 \phi^2-\frac{\delta \lambda}{4!} \phi^4 a^d - \frac{\delta\beta}{3!} \phi^3 a^d - \alpha \phi a^d  \right] 
\label{y4}
\end{eqnarray}
Note that the above quartic plus cubic potential (with $\lambda\,{\ensuremath >}\,0$) is the most general renormalisable self interaction in four spacetime dimensions.

The mode functions for a free massless scalar field $(\Box \phi=0)$ reads
$$\phi_k(x) =\frac{H}{\sqrt{2k^3}} \left(1+ik\eta \right) e^{-ik\eta+i\vec{k}\cdot \vec{x}} \qquad (k\equiv |\vec{k}|)$$
along with its complex conjugation. With these mode functions, the one can make the canonical quantisation on some initial hypersurface. We take this initial surface to be at $\eta =-1/H$ and also for the initial sub-Hubble modes, $k/H \gg 1$. The corresponding mode functions are known as the Bunch-Davies vacuum~\cite{Chernikov:1968zm, Bunch:1978yq}.   However, it is easy to check that these mode functions do not remain normalised afterwards. Accordingly, there is not de Sitter invariant two point function in the de Sitter spacetime~\cite{Allen:1985ux, Allen}.  \\   

\noindent
The propagator for a massless and minimally coupled scalar field with respect to the initial Bunch-Davies vacuum state reads, e.g.~\cite{Brunier:2004sb},
\be
i\Delta(x,x')= A(x,x')+ B(x,x')+ C(x,x')
\label{props1}
\ee
where 
\begin{eqnarray}
&&A(x,x') = \frac{H^{2-\e} \Gamma(1-\e/2)}{4\pi^{2-\e/2}}\frac{1}{y^{1-\e/2}} \nonumber\\
&&B(x,x') =  \frac{H^{2-\e} }{(4\pi)^{2-\e/2}}\left[-\frac{2\Gamma(3-\e)}{\e}\left(\frac{y}{4} \right)^{\e/2}+ \frac{2\Gamma(3-\e)}{\e\,\Gamma(2-\e/2)} + \frac{2\Gamma(3-\e)}{\Gamma(2-\e/2)}\ln (aa')\right]\nonumber\\
&&C(x,x')= \frac{H^{2-\e} }{(4\pi)^{2-\e/2}} \sum_{n=1}^{\infty} \left[\frac{\Gamma(3-\e+n)}{n\Gamma(2-\e/2+n)}\left(\frac{y}{4} \right)^n- \frac{\Gamma(3-\e/2+n)}{(n+\e/2)\Gamma(2+n)}\left(\frac{y}{4} \right)^{n+\e/2}  \right]
\label{props2}
\end{eqnarray}
where the de Sitter invariant interval is given by
\be
y(x,x')= aa'H^2 \Delta x^2 = aa'H^2 \left[ |\vec{x}-\vec{x'}|^2- (\eta-\eta')^2\right]
\label{props3}
\ee
Note that $y(x,x')$ is proportional to the invariant interval of the Minkowski spacetime, $\Delta x^2$, owing to the former's conformal flatness. 

In general in a non-equilibrium background such as the de Sitter, often the in-in or the Schwinger-Keldysh formalism needs to be used, e.g.~\cite{Weinberg, Calzetta, Hu,  Adshead}. 
There are total four propagators in this formalism, characterised by suitable four complexified  distance functions
\begin{eqnarray}
&&\Delta x^2_{++} =\left[ |\vec{x}-\vec{x'}|^2- (|\eta-\eta'|-i\e)^2\right]= (\Delta x^2_{--})^{*}\nonumber\\
&&\Delta x^2_{+-} =\left[ |\vec{x}-\vec{x'}|^2- ((\eta-\eta')+i\e)^2\right]= (\Delta x^2_{-+})^{*} \qquad (\e=0^+)
\label{props3'}
\end{eqnarray}
The first two correspond respectively to the Feynman and anti-Feynman propagators, whereas the last two correspond to the two Wightman functions. However for our present purpose, as we shall see, we  require the Feynman propagator only. The tree level propagators satisfy
$$\Box_x  i\Delta_{ss'}(x,x') = is \delta_{ss'} \delta^d(x-x')  \qquad s,s'=\pm  $$

\noindent
We have in the coincidence limit for all the four propagators
\be 
i\Delta(x,x) = \frac{H^{2-\e} \Gamma(2-\e)}{2^{2-\e} \pi^{2-\e/2} \Gamma(1-\e/2)}\left(\frac{1}{\e}+\ln a  \right)
\label{y6}
\ee
$i\Delta(x,x)$ can appear, for example, in the one loop ${\cal O}(\lambda)$ bubble self energy or one loop ${\cal O}(\beta)$ tadpole. Accordingly, the divergence can be canceled respectively by  the one loop mass renormalisation and the one loop tadpole counterterms,
\be 
 \delta m_{\lambda}^2 = - \frac{\lambda H^{2-\e} \Gamma(2-\e)}{2^{3-\e} \pi^{2-\e/2} \Gamma(1-\e/2)\e}  \qquad \qquad  \alpha= -\frac{\beta H^{2-\e} \Gamma(2-\e)}{2^{3-\e}\pi^{2-\e/2}\Gamma(1-\e/2)\e} 
\label{y7}
\ee

We also note for our purpose the square of the Feynman propagator~\cite{Brunier:2004sb}
\begin{eqnarray}
i\Delta_{++}^2(x,x')=&& -\frac{i\mu^{-\e} a'^{-4+2\e}\Gamma(1-\e/2)}{2^3\pi^{2-\e/2}\e (1-\e)}\delta^d(x-x') \,+ \frac{H^4}{2^6 \pi^4} \ln^2 \frac{\sqrt{e}H^2 \Delta x^2_{++}}{4}\nonumber\\
&& - \frac{H^2 (aa')^{-1}}{2^4\pi^4} \frac{\ln \frac{\sqrt{e}H^2 \Delta x^2_{++}}{4}}{\Delta x^2_{++}} - \frac{(aa')^{-2}}{2^6\pi^4} \p^2 \frac{\ln \mu^2\Delta x_{++}^2}{\Delta x_{++}^2} \nonumber\\
&&\equiv-\frac{i\mu^{-\e} a'^{-4+2\e}\Gamma(1-\e/2)}{2^3\pi^{2-\e/2}\e (1-\e)}\delta^d(x-x') \,+\Phi^{++}_{\rm NL} (x,x')
\label{e51'}
\end{eqnarray} 
where we have abbreviated all the non-local terms as $\Phi^{++}_{\rm NL} (x,x')$, and $\mu$ stands for an arbitrary renormalisation scale. The square of the anti-Feynman propagator is just the complex conjugation of the above. On the other hand for the mixed kind of propagators, the local term containing the $\delta$-function will be absent. \ref{e51'} can appear, for example, in the one loop ${\cal O}(\beta^2)$ self energy contribution. The divergence can be absorbed by the  mass renormalisation counterterm 
\begin{eqnarray}
\delta m_{\beta}^2= \frac{\beta^2 \mu^{-\e} \Gamma(1-\e/2)}{2^4 \pi^{2-\e/2} \e (1-\e)}
\label{e51'add}
\end{eqnarray} 
With these ingredients, we are ready to go into computing the loop corrections to the quartic and cubic vertex functions. 

%%%%%%%
\section{Loop correction to the four point vertex}\label{s3}
%%%%%
\subsection{One loop}\label{s31}
\begin{figure}[htbp]
    \centering
    \includegraphics[width=13cm, height=5.0cm]{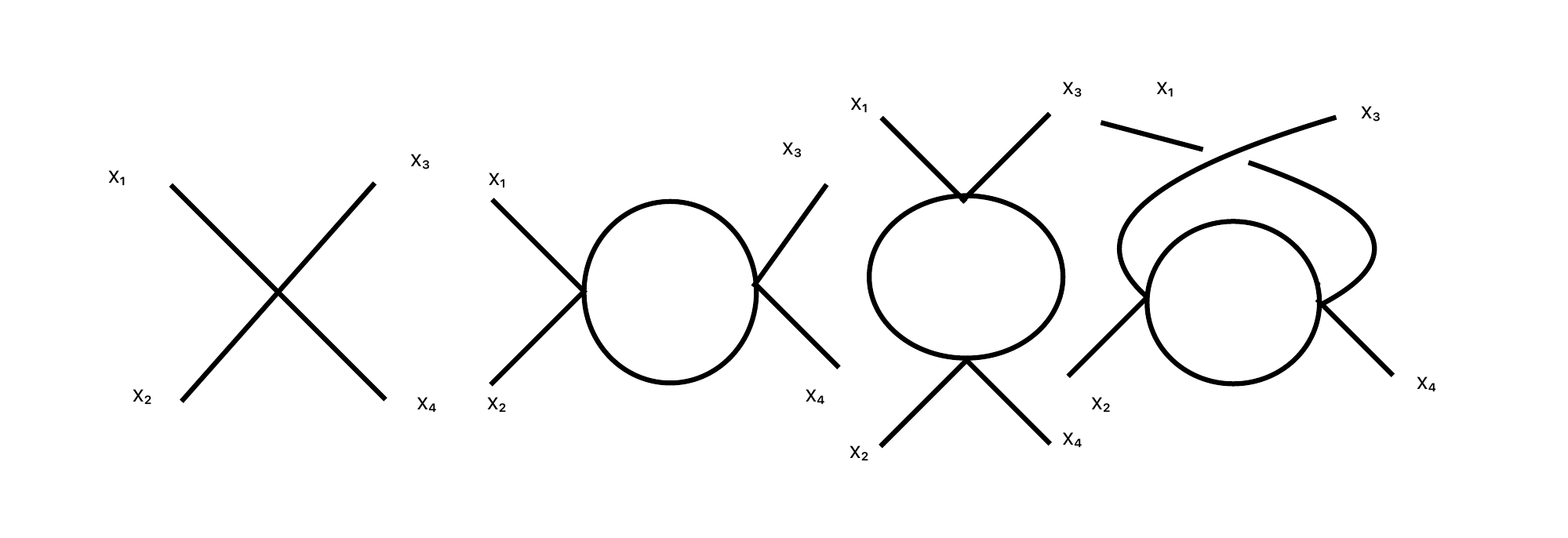}
    \caption{\small \it Quartic vertex at tree and one loop level. The three one loop diagrams correspond to the three channels, $s$, $t$ and $u$. The one loop vertex is renormalised by one loop vertex renormalisation counterterm. For each channel, the counterterm is equal. See main text for discussion. }
    \label{4-1l}
\end{figure}

%%%%%%%%%
%%%%%%%%%%%%%%%%%%%%%%%%%%%%%%%%%%%%%%%%%%%%%%%%%%%%%%%%%%%%%%%%%%%%%%%%%
The quartic vertex function at tree and one loop level is depicted in \ref{4-1l}, e.g.~\cite{Peskin}. Since we are only interested about the purely local part of the vertex functions, and the mixed kind of propagators do not have any purely local part (\ref{s2}), we shall deal only with the Feynman propagators.  

The tree level amputated $4$-point vertex can simply be read from \ref{y4},
\begin{eqnarray}
&&-iV_+(x_1,x_2,x_3,x_4)\big \vert_{\rm Tree}= -i(\lambda+\delta \lambda)a_1^d \delta^d(x_1-x_2)\delta^d(x_1-x_3)\delta^d(x_1-x_4)
\label{v1}
\end{eqnarray}

The one loop correction to the $4$-point vertex was computed earlier in~\cite{Brunier:2004sb}
\begin{eqnarray}
&&-iV_+(x_1,x_2,x_3,x_4)\big \vert_{\rm 1\,loop}= -\frac{\lambda^2}{2} \left[(a_1a_2)^d i\Delta_{++}^2(x_1,x_2) \delta^d(x_1-x_4)\delta^d(x_2-x_3)\right. \nonumber\\&&\left.+ (a_1a_3)^d i\Delta_{++}^2(x_1,x_3) \delta^d(x_1-x_2)\delta^d(x_3-x_4)+ (a_1a_4)^d i\Delta_{++}^2(x_1,x_4) \delta^d(x_1-x_3)\delta^d(x_2-x_4) \right]
\label{v2}
\end{eqnarray}

Substituting now  \ref{e51'} into the above expression, we have after a little algebra
\begin{eqnarray}
&&-iV_+(x_1,x_2,x_3,x_4)\big \vert_{\rm 1\,loop}= \frac{3i\mu^{-\e} \lambda^2}{2^4 \pi^{2-\e/2}}\frac{\Gamma(1-\e/2)}{(1-\e)\e} a_1^{4-\e+\e} \delta^d(x_1-x_2) \delta^d(x_1-x_3) \delta^d(x_1-x_4) \nonumber\\&&-\frac{\lambda^2}{2} \left[(a_1a_2)^4 \Phi_{\rm NL}^{++}(x_1,x_2) \delta^4(x_1-x_4)\delta^4(x_2-x_3)\right. \nonumber\\&&\left.+ (a_1a_3)^4 \Phi_{\rm NL}^{++}(x_1,x_3) \delta^4(x_1-x_2)\delta^4(x_3-x_4)+ (a_1a_4)^4 \Phi_{\rm NL}^{++}(x_1,x_4) \delta^4(x_1-x_3)\delta^4(x_2-x_4) \right]\nonumber\\
&& = \frac{3i\mu^{-\e} \lambda^2}{2^4 \pi^{2-\e/2}}\frac{\Gamma(1-\e/2)}{(1-\e)\e} a_1^d \delta^d(x_1-x_2) \delta^d(x_1-x_3) \delta^d(x_1-x_4) + \frac{3i\lambda^2}{2^4 \pi^{2}} a_1^4 \ln a_1 \delta^4(x_1-x_2) \delta^4(x_1-x_3) \delta^4(x_1-x_4)\nonumber\\&&-\frac{\lambda^2}{2} \left[(a_1a_2)^4 \Phi_{\rm NL}^{++}(x_1,x_2) \delta^4(x_1-x_4)\delta^4(x_2-x_3)\right. \nonumber\\&&\left.+ (a_1a_3)^4 \Phi_{\rm NL}^{++}(x_1,x_3) \delta^4(x_1-x_2)\delta^4(x_3-x_4)+ (a_1a_4)^4 \Phi_{\rm NL}^{++}(x_1,x_4) \delta^4(x_1-x_3)\delta^4(x_2-x_4) \right]
\label{v3}
\end{eqnarray}

Comparing \ref{v1} and \ref{v3}, we identify the one loop vertex counterterm
\begin{eqnarray}
\delta \lambda^{(1)} = \delta \lambda^{(1)}_s+\delta \lambda^{(1)}_t+\delta \lambda^{(1)}_u=  \frac{3\mu^{-\e} \lambda^2}{2^4 \pi^{2-\e/2}}\frac{\Gamma(1-\e/2)}{(1-\e)\e} 
\label{v4}
\end{eqnarray}
where the suffixes $s$, $t$ and $u$ represent the three Mandelstam channels with $\delta \lambda^{(1)}_s=\delta \lambda^{(1)}_t=\delta \lambda^{(1)}_u$, so that we have the renormalised result at one loop
\begin{eqnarray}
&&-iV_+(x_1,x_2,x_3,x_4)\big \vert_{\rm 1\,loop,\,Ren.}= \frac{3i\lambda^2}{2^4 \pi^{2}} a_1^4 \ln a_1 \delta^4(x_1-x_2) \delta^4(x_1-x_3) \delta^4(x_1-x_4)\nonumber\\&&-\frac{\lambda^2}{2} \left[(a_1a_2)^4 \Phi_{\rm NL}^{++}(x_1,x_2) \delta^4(x_1-x_4)\delta^4(x_2-x_3)\right. \nonumber\\&&\left.+ (a_1a_3)^4 \Phi_{\rm NL}^{++}(x_1,x_3) \delta^4(x_1-x_2)\delta^4(x_3-x_4)+ (a_1a_4)^4 \Phi_{\rm NL}^{++}(x_1,x_4) \delta^4(x_1-x_3)\delta^4(x_2-x_4) \right]
\label{v4'}
\end{eqnarray}
Setting the scale factor to unity reproduces the result of the flat spacetime, in which case no secular effect can be present. Thus the purely local and renormalised vertex function found here can have no $\Lambda=0$ analogue. Let us now compute the two loop corrected vertex function.

%%%%%%%
\subsection{Two loop}\label{s32}
%%%
%
\begin{figure}[htbp]
    \centering
    \includegraphics[width=10cm, height=8.5cm]{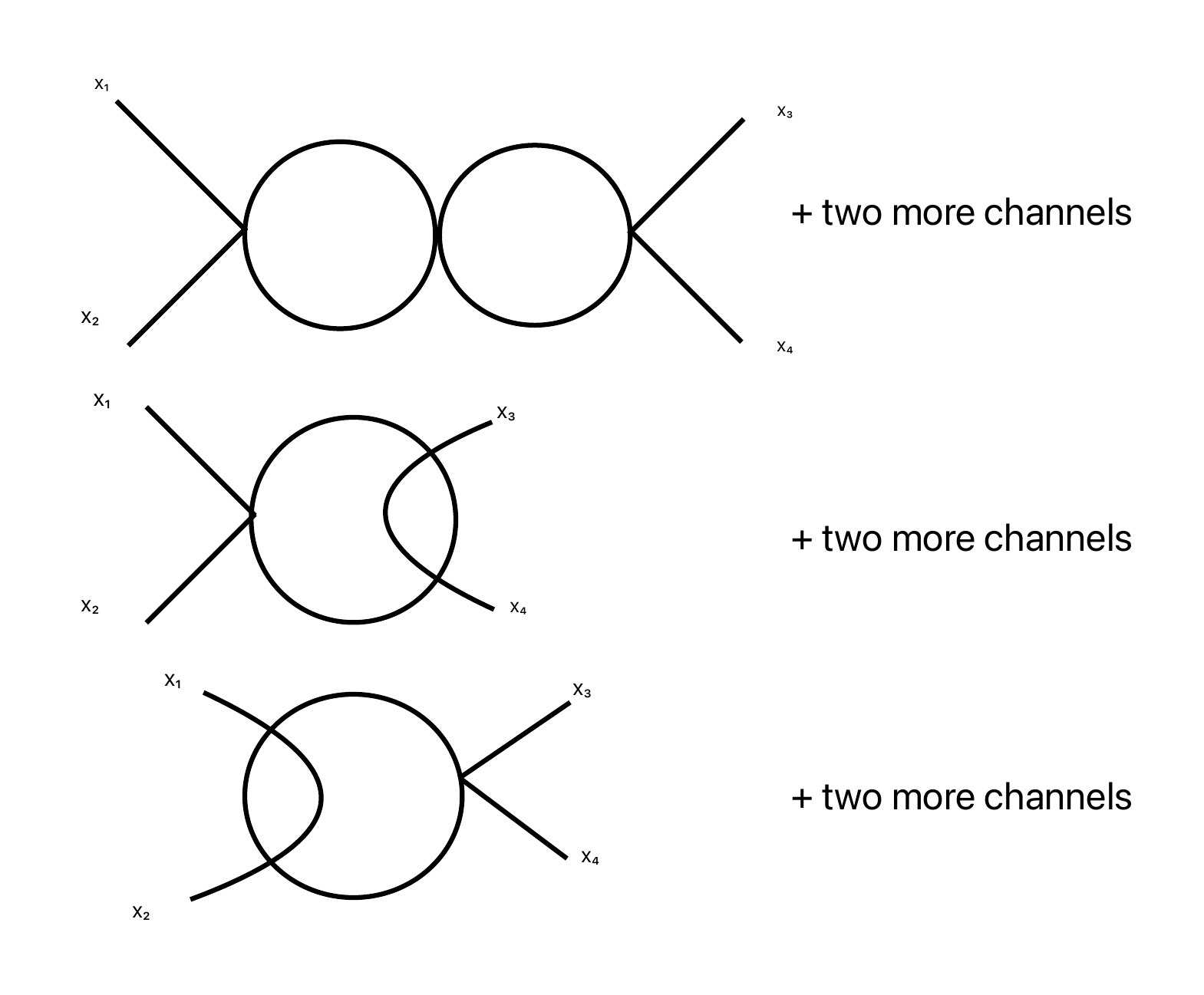}
    \caption{\small \it Quartic vertex at two loop due to quartic self interaction. A diagram in any row has three channels, e.g.~\cite{Peskin}. We have not considered the two loop diagrams at ${\cal O}(\lambda^2\beta^2)$  containing internal cubic vertices, as they do not give any purely local contribution to the vertex functions.  Likewise, the first diagram of \ref{fA1} does not also yield any purely local contributions.   See main text for discussion.   }
    \label{4-2l}
\end{figure}
%
%%%%%%%%%%%%%%%%%%%%%%%%%%%%%%%%%%%%%%%%%%%

The relevant Feynman diagrams for two loop correction to the vertex function due to the quartic self interaction are depicted in \ref{4-2l}. There is another two loop ${\cal O}(\lambda^3)$ diagram (the first of \ref{fA1}), which we have not considered here, as it does not yield any purely local correction to the four point vertex function. Let us now compute the  first row of \ref{4-2l}, 
\begin{eqnarray}
&&-iV_+(x_1,x_2,x_3,x_4)\big \vert_{\rm 2\,loop}^{(1)}= \frac{i\lambda^3 a_1^d}{2^2} \int d^d x a^d \left[a_3^d\, i\Delta_{++}^2(x_1,x)i\Delta_{++}^2(x,x_3) \delta^d(x_1-x_2)\delta^d(x_3-x_4)\right. \nonumber\\&&\left.+ a_4^d i\Delta_{++}^2(x_1,x) i\Delta_{++}^2(x,x_2)\delta^d(x_1-x_3)\delta^d(x_2-x_4)+ a_2^d i\Delta_{++}^2(x_1,x) i\Delta_{++}^2(x,x_3)\delta^d(x_1-x_4)\delta^d(x_2-x_3) \right]
\label{v5}
\end{eqnarray}

Substituting as earlier \ref{e51'} into \ref{v5}, we have 
\begin{eqnarray}
&&-iV_+(x_1,x_2,x_3,x_4)\big \vert_{\rm 2\,loop}^{(1)}
%=- \frac{3i\lambda^3 \mu^{-2\e} \Gamma^2(1-\e/2)}{2^8\pi^{4-\e} (1-\e)^2 \e^2} a_1^d a_1^{2\e}\delta^d(x_1-x_2) \delta^d(x_1-x_3) \delta^d(x_1-x_4)
=- \frac{3i\lambda^3 \mu^{-2\e} \Gamma^2(1-\e/2)}{2^8\pi^{4-\e} (1-\e)^2 } a_1^d \delta^d(x_1-x_2) \delta^d(x_1-x_3) \delta^d(x_1-x_4)\left( \frac{1}{\e^2}+ \frac{2\ln a_1}{\e} \right)
\nonumber\\&&
+ \frac{\mu^{-\e} \lambda^3 \Gamma(1-\e/2)}{2^4\pi^{2-\e/2} \e (1-\e)} \left[ (a_1a_3)^d \Phi_{\rm NL}^{++} (x_1,x_3) \delta^d(x_1-x_2)\delta^d(x_3-x_4)+(a_1a_4)^d \Phi_{\rm NL}^{++} (x_1,x_4) \delta^d(x_1-x_3)\delta^d(x_2-x_4) \right. \nonumber\\&&\left. +(a_1a_2)^d \Phi_{\rm NL}^{++} (x_1,x_2) \delta^d(x_1-x_4)\delta^d(x_2-x_3)   \right]
- \frac{3i\lambda^3  }{2^7\pi^{4}} a_1^4 \ln^2 a_1 \delta^4(x_1-x_2) \delta^4(x_1-x_3) \delta^4(x_1-x_4)\nonumber\\&&
+\frac{\lambda^3}{2^5\pi^2}\left[ (a_1a_3)^4 \ln (a_1a_3)\Phi_{\rm NL}^{++} (x_1,x_3) \delta^4(x_1-x_2)\delta^4(x_3-x_4)+(a_1a_4)^4 \ln (a_1a_4)\Phi_{\rm NL}^{++} (x_1,x_4) \delta^4(x_1-x_3)\delta^4(x_2-x_4) \right. \nonumber\\&&\left. +(a_1a_2)^4 \ln (a_1a_2)\Phi_{\rm NL}^{++} (x_1,x_2) \delta^4(x_1-x_4)\delta^4(x_2-x_3)   \right]\nonumber\\&&
+\frac{i\lambda^3 a_1^4}{2^2} \int d^4 x a^4 \left[a_3^4\, \Phi_{\rm NL}^{++}(x_1,x)\Phi_{\rm NL}^{++}(x,x_3) \delta^4(x_1-x_2)\delta^4(x_3-x_4)\right. \nonumber\\&&\left.+ a_4^4 \Phi_{\rm NL}^{++}(x_1,x) \Phi_{\rm NL}^{++}(x,x_2)\delta^4(x_1-x_3)\delta^4(x_2-x_4)+ a_2^4 \Phi_{\rm NL}^{++}(x_1,x) \Phi_{\rm NL}^{++}(x,x_3)\delta^4(x_1-x_4)\delta^4(x_2-x_3) \right]
\label{v6}
\end{eqnarray}

We now add with it the one loop vertex counterterm diagrams. We need to distinguish between the different channels here.  From \ref{v4}, we have 
\begin{eqnarray}
&&-iV_+(x_1,x_2,x_3,x_4)\big \vert_{\rm CT}= -\lambda \left[\delta \lambda_s(a_1a_2)^d  i\Delta_{++}^2(x_1,x_2) \delta^d(x_1-x_4)\delta^d(x_2-x_3)\right. \nonumber\\&&\left.+\delta \lambda_t  (a_1a_3)^d i\Delta_{++}^2(x_1,x_3) \delta^d(x_1-x_2)\delta^d(x_3-x_4)+ \delta \lambda_u (a_1a_4)^d i\Delta_{++}^2(x_1,x_4) \delta^d(x_1-x_3)\delta^d(x_2-x_4) \right]\nonumber\\&&
=\frac{3i\mu^{-2\e} \lambda^3 \Gamma^2(1-\e/2)}{2^7 \pi^{4-\e} (1-\e)^2}a_1^d\left[\frac{1}{ \e^2}+\frac{\ln a_1}{\e} + \frac{\ln^2 a_1}{2}\right] \delta^d(x_1-x_2)\delta^d(x_1-x_3)\delta^d(x_1-x_4)\nonumber\\&&
-\frac{\mu^{-\e} \lambda^3 \Gamma(1-\e/2)}{2^4 \pi^{2-\e/2} (1-\e)\e}\left[ (a_1a_2)^d \Phi_{\rm NL}^{++}(x_1,x_2) \delta^d(x_1-x_4)\delta^d(x_2-x_3)+ (a_1a_3)^d \Phi_{\rm NL}^{++}(x_1,x_3) \delta^d(x_1-x_2)\delta^d(x_3-x_4)\right. \nonumber\\&&\left.+(a_1a_4)^d \Phi_{\rm NL}^{++}(x_1,x_4) \delta^d(x_1-x_3)\delta^d(x_2-x_4)\right]
\label{v6'}
\end{eqnarray}

Adding the above with \ref{v6}, we have
\begin{eqnarray}
&&-iV_+(x_1,x_2,x_3,x_4)\big \vert_{\rm 2\,loop}^{(1)}=\frac{3i\mu^{-2\e} \lambda^3 \Gamma^2(1-\e/2)}{2^8 \pi^{4-\e} (1-\e)^2\e^2}a_1^d \delta^d(x_1-x_2)\delta^d(x_1-x_3)\delta^d(x_1-x_4)\nonumber\\&&- \frac{3i\lambda^3  }{2^8\pi^{4}} a_1^4 \ln^2 a_1 \delta^4(x_1-x_2) \delta^4(x_1-x_3) \delta^4(x_1-x_4)+\frac{\lambda^3}{2^5\pi^2}\left[ (a_1a_3)^4 \ln (a_1a_3)\,\Phi_{\rm NL}^{++} (x_1,x_3) \delta^4(x_1-x_2)\delta^4(x_3-x_4)\right. \nonumber\\&&\left.+(a_1a_4)^4 \ln (a_1a_4)\,\Phi_{\rm NL}^{++} (x_1,x_4) \delta^4(x_1-x_3)\delta^4(x_2-x_4)  +(a_1a_2)^4 \ln (a_1a_2)\,\Phi_{\rm NL}^{++} (x_1,x_2) \delta^4(x_1-x_4)\delta^4(x_2-x_3)   \right]\nonumber\\
&&+\frac{i\lambda^3 a_1^4}{2^2} \int d^4 x a^4 \left[a_3^4\, \Phi_{\rm NL}^{++}(x_1,x)\Phi_{\rm NL}^{++}(x,x_3) \delta^4(x_1-x_2)\delta^4(x_3-x_4)\right. \nonumber\\&&\left.+ a_4^4 \Phi_{\rm NL}^{++}(x_1,x) \Phi_{\rm NL}^{++}(x,x_2)\delta^4(x_1-x_3)\delta^4(x_2-x_4)+ a_2^4 \Phi_{\rm NL}^{++}(x_1,x) \Phi_{\rm NL}^{++}(x,x_3)\delta^4(x_1-x_4)\delta^4(x_2-x_3) \right]
\label{v6''}
\end{eqnarray}
We shall renormalise the divergence in the first line after we compute the two other two loop vertex diagrams of \ref{4-2l}. Note also that we have additional divergent terms due to the secular logarithms compared to the flat spacetime in \ref{v6}. Those additional divergences are taken care of by the additional divergent terms of the counterterm contribution, \ref{v6'}.\\

\noindent
The second row of \ref{4-2l} reads
\begin{eqnarray}
&&-iV_+(x_1,x_2,x_3,x_4)\big \vert_{\rm 2\,loop}^{(2)}=\frac{i\lambda^3 a_1^d}{2}\left[ (a_2a_3)^d i\Delta_{++}(x_1,x_3)i\Delta_{++}(x_1,x_2)i\Delta^2_{++}(x_2,x_3)\delta^d(x_1-x_4)\right. \nonumber\\&&\left.+(a_3a_4)^d i\Delta_{++}(x_1,x_3)i\Delta_{++}(x_1,x_4)i\Delta^2_{++}(x_3,x_4)\delta^d(x_1-x_2)+(a_4a_2)^di\Delta_{++}(x_1,x_2)i\Delta_{++}(x_1,x_4)i\Delta^2_{++}(x_2,x_4)\delta^d(x_1-x_3)\right]\nonumber\\
\label{v7}
\end{eqnarray}
Substituting \ref{e51'} into the above expression, we obtain  
\begin{eqnarray}
&&-iV_+(x_1,x_2,x_3,x_4)\big \vert_{\rm 2\,loop,\,loc}^{(2)}=- \frac{3i\lambda^3 \mu^{-2\e} \Gamma^2(1-\e/2)}{2^7\pi^{4-\e} (1-\e)^2 } a_1^d \delta^d(x_1-x_2) \delta^d(x_1-x_3) \delta^d(x_1-x_4)\left( \frac{1}{\e^2}+ \frac{2\ln a_1}{\e} \right)\nonumber\\&&- \frac{3i\lambda^3 }{2^6\pi^{4} } a_1^4 \, \ln^2a_1\delta^4(x_1-x_2) \delta^4(x_1-x_3) \delta^4(x_1-x_4) +\frac{\mu^{-\e} \lambda^3 \Gamma(1-\e/2)}{2^3\pi^{2-\e/2}\e(1-\e)} \left[(a_1a_3)^d \Phi_{\rm NL}^{++} (x_1,x_3) \delta^d(x_2-x_3) \delta^d(x_1-x_4) \right. \nonumber\\&&\left. + (a_1a_4)^d \Phi_{\rm NL}^{++} (x_1,x_4) \delta^d(x_1-x_2) \delta^d(x_3-x_4)+(a_1a_2)^d \Phi_{\rm NL}^{++} (x_1,x_2) \delta^d(x_1-x_3) \delta^d(x_2-x_4)    \right]\nonumber\\
&&+ \frac{ \lambda^3 }{2^3\pi^{2}} \left[(a_1a_3)^4\ln a_3\, \Phi_{\rm NL}^{++} (x_1,x_3) \delta^4(x_2-x_3) \delta^4(x_1-x_4) \right. \nonumber\\&&\left. + (a_1a_4)^4 \ln a_4\, \Phi_{\rm NL}^{++} (x_1,x_4) \delta^4(x_1-x_2) \delta^4(x_3-x_4)+(a_1a_2)^4 \ln a_2\, \Phi_{\rm NL}^{++} (x_1,x_2) \delta^4(x_1-x_3) \delta^4(x_2-x_4)    \right]\nonumber\\
&&+ \frac{i\lambda^3 a_1^4}{2}\left[ (a_2a_3)^4 \Phi_{\rm NL}^{++} (x_1,x_3)\Phi_{\rm NL}^{++} (x_2,x_3)\delta^4(x_1-x_4)+(a_3a_4)^4 \Phi_{\rm NL}^{++} (x_1,x_3)\Phi_{\rm NL}^{++} (x_3,x_4)\delta^4(x_1-x_2)\right. \nonumber\\&&\left. +  (a_4a_1)^4 \Phi_{\rm NL}^{++} (x_2,x_1)\Phi_{\rm NL}^{++} (x_1,x_4)\delta^4(x_2-x_3)\right]
\label{v7'}
\end{eqnarray}
A clarification is in order here. Since each of the channels contains two $\delta$-functions, the channels are obtained only when  we encounter a square of the propagator. Thus in order to obtain the full channel wise contributions,  for example for the first term on the right hand side of  \ref{v7} we must take into account the effect when $x_2\to x_3$, in the product $ i\Delta_{++}(x_1,x_3)i\Delta_{++}(x_1,x_2)$. Ignoring this still gives a channel structure, however the respective contributions become half, which is incomplete.  As in the flat 
spacetime~\cite{Peskin}, the above diagram is renormalised by adding with it the $(s+t)$-counterterm contributions, \ref{v4}, plus two more channels (corresponding to $(t+u)$ and $(u+s)$ counterterms), which is just twice of \ref{v6'}. Adding this with \ref{v7'}, we obtain the contribution free from any $\ln a /\e$-type divergences,
\begin{eqnarray}
&&-iV_+(x_1,x_2,x_3,x_4)\big \vert_{\rm 2\,loop}^{(2)}= \frac{3i\lambda^3 \mu^{-2\e} \Gamma^2(1-\e/2)}{2^7\pi^{4-\e} (1-\e)^2\e^2 } a_1^d \delta^d(x_1-x_2) \delta^d(x_1-x_3) \delta^d(x_1-x_4) \nonumber\\&&- \frac{3i\lambda^3 }{2^7\pi^{4} } a_1^4 \, \ln^2a_1\delta^4(x_1-x_2) \delta^4(x_1-x_3) \delta^4(x_1-x_4) + \frac{ \lambda^3 }{2^3\pi^{2}} \left[(a_1a_3)^4\ln a_3\, \Phi_{\rm NL}^{++} (x_1,x_3) \delta^4(x_2-x_3) \delta^4(x_1-x_4) \right. \nonumber\\&&\left. + (a_1a_4)^4 \ln a_4\, \Phi_{\rm NL}^{++} (x_1,x_4) \delta^4(x_1-x_2) \delta^4(x_3-x_4)+(a_1a_2)^4 \ln a_2\, \Phi_{\rm NL}^{++} (x_1,x_2) \delta^4(x_1-x_3) \delta^4(x_2-x_4)    \right]
\nonumber\\
&&+ \frac{i\lambda^3 a_1^4}{2}\left[ (a_2a_3)^4 \Phi_{\rm NL}^{++} (x_1,x_3)\Phi_{\rm NL}^{++} (x_2,x_3)\delta^4(x_1-x_4)+(a_3a_4)^4 \Phi_{\rm NL}^{++} (x_1,x_3)\Phi_{\rm NL}^{++} (x_3,x_4)\delta^4(x_1-x_2)\right. \nonumber\\&&\left. +  (a_4a_1)^4 \Phi_{\rm NL}^{++} (x_2,x_1)\Phi_{\rm NL}^{++} (x_1,x_4)\delta^4(x_2-x_3)\right]
\label{v7''}
\end{eqnarray}
\\

\noindent
Finally, the third row of \ref{4-2l} reads
\begin{eqnarray}
&&-iV_+(x_1,x_2,x_3,x_4)\big \vert_{\rm 2\,loop}^{(3)}=\frac{i\lambda^3 a_2^d}{2}\left[ (a_1a_3)^d i\Delta_{++}(x_2,x_3)i\Delta_{++}(x_2,x_1)i\Delta^2_{++}(x_1,x_3)\delta^d(x_2-x_4)\right. \nonumber\\&&\left.+(a_3a_4)^d i\Delta_{++}(x_2,x_3)i\Delta_{++}(x_2,x_4)i\Delta^2_{++}(x_3,x_4)\delta^d(x_2-x_1)+(a_4a_1)^di\Delta_{++}(x_2,x_1)i\Delta_{++}(x_2,x_4)i\Delta^2_{++}(x_1,x_4)\delta^d(x_2-x_3)\right]\nonumber\\
\label{v8}
\end{eqnarray}
The analogue of \ref{v7''} in this case can be derived in a likewise manner
\begin{eqnarray}
&&-iV_+(x_1,x_2,x_3,x_4)\big \vert_{\rm 2\,loop,\,loc}^{(3)}= \frac{3i\lambda^3 \mu^{-2\e} \Gamma^2(1-\e/2)}{2^7\pi^{4-\e} (1-\e)^2 \e^2} a_2^d \delta^d(x_2-x_1) \delta^d(x_2-x_3) \delta^d(x_2-x_4)\nonumber\\&&- \frac{3i\lambda^3 }{2^7\pi^{4} } a_2^4 \, \ln^2a_2\delta^4(x_2-x_1) \delta^4(x_2-x_3) \delta^4(x_2-x_4) + \frac{ \lambda^3 }{2^3\pi^{2}} \left[(a_2a_3)^4\ln a_3\, \Phi_{\rm NL} (x_2,x_3) \delta^4(x_1-x_3) \delta^4(x_2-x_4) \right. \nonumber\\&&\left. + (a_2a_4)^4 \ln a_4\, \Phi_{\rm NL} (x_2,x_4) \delta^4(x_2-x_1) \delta^4(x_3-x_4)+(a_2a_1)^4 \ln a_1\, \Phi_{\rm NL} (x_2,x_1) \delta^4(x_2-x_3) \delta^4(x_1-x_4)    \right]
\nonumber\\
&&+ \frac{i\lambda^3 a_1^4}{2}\left[ (a_2a_3)^4 \Phi_{\rm NL}^{++} (x_1,x_3)\Phi_{\rm NL}^{++} (x_2,x_3)\delta^4(x_2-x_4)+(a_3a_4)^4 \Phi_{\rm NL}^{++} (x_2,x_3)\Phi_{\rm NL}^{++} (x_3,x_4)\delta^4(x_2-x_1)\right. \nonumber\\&&\left. +  (a_4a_2)^4 \Phi_{\rm NL}^{++} (x_2,x_1)\Phi_{\rm NL}^{++} (x_1,x_4)\delta^4(x_2-x_3)\right]
\label{v9}
\end{eqnarray}
Note by comparison of the above with \ref{v7'} that these purely local parts are not really different by the virtue of the $\delta$-functions, as expected, and one can be obtained from the other by purely making the interchange $x_1 \leftrightarrow x_2$. However, we have kept them in a little bit different appearances  just to emphasise  the different topologies of the corresponding diagrams. 

Combining now \ref{v9}, \ref{v7''} and \ref{v6''}, we find the only remaining two loop divergent term for \ref{4-2l} to be 
\be
-iV_+(x_1,x_2,x_3,x_4)\big \vert_{\rm 2\,loop,\, div}^{(3)}=3\times \frac{5i\lambda^3 \mu^{-2\e} \Gamma^2(1-\e/2)}{2^8\pi^{4-\e} (1-\e)^2 \e^2} a_1^d \delta^d(x_1-x_2) \delta^d(x_1-x_3) \delta^d(x_1-x_4)
\label{v10}
\ee
which leads to the two loop quartic vertex counterterm,
\be
\delta \lambda^{(2)}= \delta \lambda_s^{(2)} +\delta \lambda_t^{(2)}+\delta \lambda^{(2)}_u, \qquad 
\delta \lambda_s^{(2)} =\delta \lambda_t^{(2)}=\delta \lambda^{(2)}_u= \frac{5\lambda^3 \mu^{-2\e} \Gamma^2(1-\e/2)}{2^8\pi^{4-\e} (1-\e)^2 \e^2} 
\label{v11}
\ee
which completely renormalises the addition of \ref{v9}, \ref{v7''} and \ref{v6''}. Since we are  interested only about the purely local vertex function, we pick up the term 
\be
-iV_+(x_1,x_2,x_3,x_4)\big \vert_{\rm 2\,loop,\,loc,\,Ren}=- \frac{15i\lambda^3 }{2^8\pi^{4} } a_1^4 \ln^2 a_1 \delta^d(x_1-x_2) \delta^d(x_1-x_3) \delta^d(x_1-x_4)
\label{v12}
\ee
\\

\noindent
Combining the above with \ref{v4'} and \ref{v1}, we have the purely local, renormalised contribution to the $4$-point vertex function up to two loop
\be
-iV_+(x_1,x_2,x_3,x_4)\big \vert_{\rm loc,\,Ren}=-\left[ \lambda - \frac{3\lambda^2}{2^4 \pi^{2}}  \ln a_1 + \frac{15\lambda^3 }{2^8\pi^{4} }  \ln^2 a_1\right]ia_1^4\delta^4(x_1-x_2) \delta^4(x_1-x_3) \delta^4(x_1-x_4)+{\cal O}(\lambda^4)
\label{v13}
\ee
which leads to a natural decomposition,  
\be
-iV_+(x_1,x_2,x_3,x_4)\big \vert_{\rm loc,\,Ren}\equiv -i a_1^4 \delta^4(x_1-x_2) \delta^4(x_1-x_3) \delta^4(x_1-x_4) \times \lambda_{\rm eff,\,loc}(a_1)
\label{v14}
\ee
where the spatially homogeneous local vertex function $\lambda_{\rm eff,\,loc}(a_1)$ contains the secular logarithms
\be 
\lambda_{\rm eff,\,loc}(a):= \lambda -\frac{3\lambda^2}{2^4 \pi^{2}}  \ln a + \frac{15\lambda^3 }{2^8\pi^{4} }  \ln^2 a + {\cal O}(\lambda^4)
\label{v15}
\ee
The appearances of the secular logarithm terms clearly indicates the breakdown of the perturbation theory at sufficiently late times. This also shows the necessity of resummation, which we wish to  pursue in \ref{s5}.

In \ref{A}, we have sketched the computation of the other ${\cal O}(\lambda^3)$ diagram, which does not yield any purely local contribution to the vertex function. Note also that contribution from such diagram is vanishing in flat spacetime, once we renormalise it. However in the de Sitter spacetime, as we have discussed in \ref{A}, the renormalised expression is non-vanishing, owing to the secular logarithm. We now wish to do similar computations for the cubic vertex function in the presence of a quartic self interaction. The loop correction due to the cubic self interaction alone does not yield any purely local contribution to the 3-point vertex function.

%%%%%%
\section{Loop correction to the cubic vertex function}\label{s4}
%%%%%%
%
\begin{figure}[htbp]
    \centering
    \includegraphics[width=12cm, height=9cm]{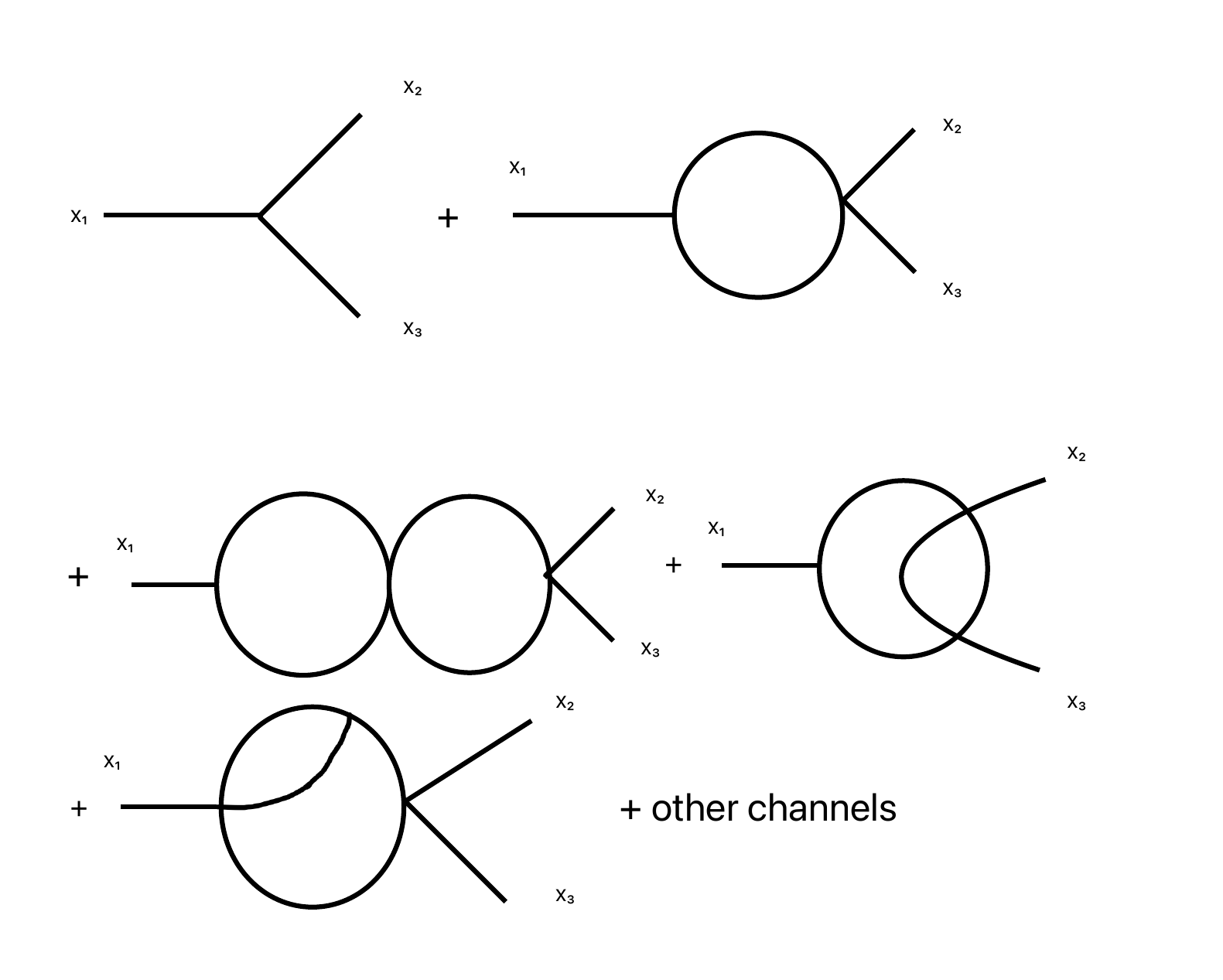}
    \caption{\small \it Cubic vertex function up to two loop for a cubic plus quartic self interaction. We have not considered corrections due to the cubic self interaction itself, for any such diagram does not yield any purely local contribution. We have also discussed another diagram in \ref{fA1} (\ref{A}) in the presence of quartic plus cubic coupling. Although this particular diagram does not produce any purely local contribution, it has a non-vanishing renormalised (non-local) contribution by the virtue of secular logarithms. Such non-vanishing contribution is not possible in the flat spacetime. See main text for discussion.}
    \label{3-2l}
\end{figure}
The cubic vertex function at tree, one loop and two loop level has been depicted in \ref{3-2l}, respectively at ${\cal O}(\beta)$, ${\cal O}(\lambda \beta)$ and ${\cal O}(\lambda^2 \beta)$ and have discarded the diagrams which do not make any purely local contribution to the cubic vertex function. We have considered another ${\cal O}(\lambda^2 \beta)$ diagram in \ref{A}, \ref{fA1}. Although this diagram does not contribute to purely local vertex functions, we have addressed it in order emphasise that it has a non-vanishing renormalised expression via the secular logarithm. In the flat spacetime however, this diagram vanishes after renormalisation.\\

\noindent
The tree level cubic vertex is given by the first of \ref{3-2l} and \ref{y4},
\be
-i V_+(x_1,x_2,x_3)= -i(\beta+\delta\beta) a_1^d \delta^d(x_1-x_2)\delta^d(x_1-x_3)
\label{cv1}
\ee
%

%%%%%
\subsection{One loop}
%%%%%

The one loop correction to the cubic vertex  due to the quartic self interaction, given by the second of \ref{3-2l}, reads
\begin{eqnarray}
&&-i V_+(x_1,x_2,x_3)\vert_{\rm 1\,loop}= -\frac{\lambda \beta}{2} \left[(a_1a_3)^d i\Delta_{++}^2(x_1,x_3) \delta^d(x_1-x_2) \right. \nonumber\\&&\left.+(a_1a_2)^d i\Delta_{++}^2(x_1,x_2)\delta^d(x_1-x_3)+ (a_1a_3)^d i\Delta_{++}^2(x_3,x_1)\delta^d(x_2-x_3) \right]\nonumber\\&&=\frac{3i\mu^{-\e}\lambda \beta \Gamma(1-\e/2)}{2^4\pi^{2-\e/2} \e(1-\e)}a_1^d  \delta^d(x_1-x_3)\delta^d(x_1-x_2)+ \frac{3i\lambda \beta }{2^4\pi^{2} }a_1^4 \ln a_1  \delta^4(x_1-x_3)\delta^4(x_1-x_2)\nonumber\\&&-\frac{\lambda \beta}{2} \left[(a_1a_3)^4 \Phi_{\rm NL}^{++}(x_1,x_3) \delta^4(x_1-x_2)+ (a_1a_2)^4 \Phi_{\rm NL}^{++}(x_1,x_2)\delta^d(x_1-x_3)+ (a_1a_3)^4 \Phi_{\rm NL}^{++}(x_3,x_1)\delta^d(x_2-x_3) \right] \hskip .2cm
\label{cv2}
\end{eqnarray}
where as earlier, we have substituted \ref{e51'}.

The one loop cubic vertex counterterm for  different channels can then be read off from \ref{cv1},
\begin{eqnarray}
\delta \beta_{\rm 1\, loop}=  \left( \delta \beta_s+ \delta \beta_t+ \delta \beta_u \right)_{\rm 1\, loop}= \frac{3\mu^{-\e}\lambda \beta \Gamma(1-\e/2)}{2^4\pi^{2-\e/2} \e(1-\e)}
\label{cv3}
\end{eqnarray}
Note that as of the pure quartic self interaction discussed in the preceding section, the different channels' contribution to the counterterm is the same. We now note down for our present purpose the purely local renormalised cubic vertex function at one loop, 
\begin{eqnarray}
-i V_+(x_1,x_2,x_3)\vert_{\rm 1\,loop,\,loc, \,Ren.}=\frac{3i\lambda \beta }{2^4\pi^{2}}a_1^4 \ln a_1 \delta^4(x_1-x_3)\delta^4(x_1-x_2)
\label{cv3'}
\end{eqnarray}
As we have stated earlier, the ${\cal O}(\beta^3)$ one loop contribution has no purely local part, and hence will not be considered here for our present purpose. 

%%%%%%
\subsection{Two loop}
%%%%%

The two loop diagrams for the cubic vertex function which contribute to the purely local vertex function are given by the second and third row of \ref{3-2l}. Diagram of each category contains three channels. The computations will be similar to that of the quartic vertex, discussed in the preceding section.  The first two loop diagram of \ref{3-2l} reads
\begin{eqnarray}
&&-i V_+(x_1,x_2,x_3)\vert_{\rm 2\,loop}^{(1)}= \frac{i\lambda^2 \beta}{2^2} \left[(a_1a_2)^d \int d^d x\, a^d\, i\Delta^2_{++}(x_1,x) i\Delta^2_{++}(x,x_2)\,\delta^d (x_2-x_3)\right. \nonumber\\&&\left.
+(a_1a_3)^d  \int d^d x\, a^d\, i\Delta^2_{++}(x_1,x) i\Delta^2_{++}(x,x_2)\,\delta^d (x_1-x_3)+ (a_2a_3)^d  \int d^d x\, a^d\, i\Delta^2_{++}(x_1,x) i\Delta^2_{++}(x,x_3)\,\delta^d (x_1-x_2) \right]\nonumber\\
&&= \frac{i\lambda^2 \beta}{2^2}\left[-\frac{3\mu^{-2\e}\Gamma^2(1-\e/2)}{2^6\pi^{4-\e}  (1-\e)^2} a_1^d \delta^d(x_1-x_2)\delta^d(x_1-x_3)\left(\frac{1}{\e^2}+\frac{2\ln a_1}{\e}+ 2\ln^2 a_1 \right) \right. \nonumber\\ &&\left.  -\frac{i\mu^{-\e} \Gamma(1-\e/2)}{2^2\pi^{2-\e/2} \e(1-\e)} \left((a_1a_2)^d \Phi^{++}_{\rm NL}(x_1,x_2)\delta^d (x_2-x_3) + (a_1a_3)^d \Phi^{++}_{\rm NL}(x_1,x_2)\delta^d (x_1-x_3)+(a_2a_3)^d \Phi^{++}_{\rm NL}(x_2,x_3)\delta^d (x_1-x_2) \right) \right. \nonumber\\ &&\left.-\frac{i (a_1a_2)^4}{2^3\pi^2}\left(\ln (a_1a_2) \Phi^{++}_{\rm NL}(x_1,x_2)\delta^4 (x_2-x_3)+\ln (a_3a_2) \Phi^{++}_{\rm NL}(x_3,x_2)\delta^4 (x_1-x_3) + \ln (a_1a_3) \Phi^{++}_{\rm NL}(x_1,x_3)\delta^4 (x_1-x_2) \right) \right. \nonumber\\ &&\left.+ \left((a_1a_2)^4 \int d^4 x\, a^4 \Phi^{++}_{\rm NL}(x_1,x)\Phi^{++}_{\rm NL}(x,x_2)\delta^4 (x_2-x_3) + (a_1a_3)^4 \int d^4 x \, a^4\Phi^{++}_{\rm NL}(x_1,x)\Phi^{++}_{\rm NL}(x,x_2)\delta^4 (x_1-x_3)    \right. \right.\nonumber\\ &&\left.\left.+ (a_2a_3)^4 \int d^4 x \,a^4\Phi^{++}_{\rm NL}(x_1,x)\Phi^{++}_{\rm NL}(x,x_3)\delta^4 (x_1-x_2)\right)  \right]
\label{cv4}
\end{eqnarray}

We add with the above the one loop counterterm contributions for different channels from \ref{v4}, \ref{cv3}, 
\begin{eqnarray}
&&-i V_+(x_1,x_2,x_3)\vert_{\rm 1\,loop\,CT}= -\frac{\beta \delta \lambda_s+\lambda \delta \beta_s }{2}(a_1a_3)^d i\Delta^2_{++}(x_1,x_3)\delta^d (x_1-x_2) \,+\, t\,\,{\rm and }  \,\,u\,\,{\rm channel~contributions}\nonumber\\
%&&=-\frac{\mu^{-\e}\beta \lambda^2 \Gamma(1-\e/2)}{2^4\pi^{2-\e/2}\e(1-\e)}(a_1a_3)^d i\Delta^2_{++}(x_1,x_3)\delta^d (x_1-x_2) \,+\, t\,\,{\rm and }  \,\,u\,\,{\rm channels}\nonumber\\
&&= \frac{3i\mu^{-2\e}\beta \lambda^2 \Gamma^2(1-\e/2)}{2^7\pi^{4-\e}(1-\e)^2}a_1^d \delta^d(x_1-x_3)\delta^d(x_1-x_2)\left(\frac{1}{\e^2}+\frac{\ln a_1}{\e}+\frac{\ln^2 a_1}{2} \right)\nonumber\\&&
-\frac{\mu^{-\e}\beta\lambda^2 \Gamma(1-\e/2)}{2^4\pi^{2-\e/2}\e(1-\e)}\left[(a_1a_2)^d \Phi^{++}_{\rm NL}(x_1,x_2)\delta^d (x_2-x_3) + (a_1a_3)^d \Phi^{++}_{\rm NL}(x_1,x_2)\delta^d (x_1-x_3)+(a_2a_3)^d \Phi^{++}_{\rm NL}(x_2,x_3)\delta^d (x_1-x_2) \right]\nonumber\\
\label{cv5}
\end{eqnarray}
so that the purely local part of \ref{cv4} is given by 
\begin{eqnarray}
-i V_+(x_1,x_2,x_3)\vert_{\rm 2\,loop\,loc}^{(1)}= -   \frac{3i\mu^{-2\e}\beta \lambda^2 \Gamma^2(1-\e/2)}{2^8\pi^{4-\e}(1-\e)^2}a_1^d \delta^d(x_1-x_3)\delta^d(x_1-x_2)\left(-\frac{1}{\e^2}+\ln^2 a_1 \right)
\label{cv5'}
\end{eqnarray}
\\

\noindent
The next two loop diagram of \ref{3-2l} reads 
\begin{eqnarray}
&&-i V_+(x_1,x_2,x_3)\vert_{\rm 2\,loop}^{(2)}= \frac{i\beta \lambda^2}{2}(a_1a_2a_3)^d \left[i\Delta_{++}(x_1,x_2)i\Delta_{++}(x_1,x_3)i\Delta^2_{++}(x_2,x_3)\right.\nonumber\\ && \left.+i\Delta_{++}(x_2,x_1)i\Delta_{++}(x_2,x_3)i\Delta^2_{++}(x_3,x_1)+i\Delta_{++}(x_3,x_1)i\Delta_{++}(x_3,x_2)i\Delta^2_{++}(x_1,x_2) \right]\nonumber\\
%&&=\frac{\mu^{-\e}\beta \lambda^2 \Gamma(1-\e/2)}{2^4 \pi^{2-\e/2}\e(1-\e)}(a_1a_2a_3)^d \left[ i\Delta_{++}^2(x_1,x_2)a_3^{-4+2\e}\delta^d(x_2-x_3)+i\Delta_{++}^2(x_2,x_1)a_1^{-4+2\e}\delta^d(x_1-x_3)\right. \nonumber\\&&\left.+i\Delta_{++}^2(x_3,x_1)a_2^{-4+2\e}\delta^d(x_1-x_2)\right]+\frac{i\beta \lambda^2}{2}(a_1a_2a_3)^d \left[i\Delta_{++}(x_1,x_2)i\Delta_{++}(x_1,x_3)\Phi_{\rm NL}^{++}(x_2,x_3)\right.\nonumber\\ && \left.+i\Delta_{++}(x_2,x_1)i\Delta_{++}(x_2,x_3)\Phi_{\rm NL}^{++}(x_3,x_1)+i\Delta_{++}(x_3,x_1)i\Delta_{++}(x_3,x_2)\Phi_{\rm NL}^{++}(x_1,x_2) \right]\nonumber\\
&&= -\frac{3i\mu^{-2\e} \beta \lambda^2 \Gamma^2(1-\e/2)}{2^7 \pi^{4-\e} (1-\e)^2}\left(\frac{1}{\e^2}+\frac{2\ln a_1}{\e}+2\ln^2 a_1 \right)+  \frac{\mu^{-\e} \beta \lambda^2 \Gamma(1-\e/2)}{2^3 \pi^{2-\e/2} \e(1-\e)}\left[(a_1a_2)^d \Phi_{\rm NL}^{++}(x_1,x_2)\delta^d(x_2-x_3)\right. \nonumber\\&&\left.
+(a_2a_3)^d \Phi_{\rm NL}^{++}(x_2,x_3)\delta^d(x_1-x_3)+(a_3a_1)^d \Phi_{\rm NL}^{++}(x_3,x_1)\delta^d(x_1-x_2)\right]+ \frac{\beta \lambda^2}{2^4 \pi^2}\left[(a_1a_2)^4\ln (a_1a_2) \Phi_{\rm NL}^{++}(x_1,x_2)\delta^4(x_2-x_3)\right. \nonumber\\&&\left.
+(a_2a_3)^4 \ln (a_2a_3)\Phi_{\rm NL}^{++}(x_2,x_3)\delta^4(x_1-x_3)+(a_3a_1)^4\ln (a_1 a_3) \Phi_{\rm NL}^{++}(x_3,x_1)\delta^4(x_1-x_2)\right]\,+\,{\rm purely~non-local~terms} \nonumber\\
\label{cv6}
\end{eqnarray}
where we have suppressed the purely non-local terms containing no $\delta$-functions, as for our present purpose they will not be necessary.  All but the ${\cal O}(\e^{-2})$ divergence of \ref{cv6} can be canceled as of \ref{v7'} (corresponding to the second or third of \ref{4-2l}), discussed in the preceding section. We have for the purely local part 
\begin{eqnarray}
-i V_+(x_1,x_2,x_3)\vert_{\rm 2\,loop\,loc}^{(2)}= -   \frac{3i\mu^{-2\e}\beta \lambda^2 \Gamma^2(1-\e/2)}{2^7\pi^{4-\e}(1-\e)^2}a_1^d \delta^d(x_1-x_3)\delta^d(x_1-x_2)\left(-\frac{1}{\e^2}+\ln^2 a_1 \right)
\label{cv6'}
\end{eqnarray}
\\

\noindent
The last two loop diagram of \ref{3-2l} for the cubic vertex reads,
\begin{eqnarray}
-i V_+(x_1,x_2,x_3)\vert_{\rm 2\,loop}^{(3)} = \frac{i\lambda^2 \beta}{2}\left[(a_1a_2)^d \int d^d x a^d i\Delta_{++}^2(x_1,x) i\Delta_{++}(x,x_2)i\Delta_{++}(x_1,x_3)\delta^d (x_2-x_3)\,+\,{\rm two~more~channels}\right], \nonumber\\
\label{cubic-add}
\end{eqnarray}
which can be evaluated as earlier to obtain the purely local contribution
\begin{eqnarray}
-i V_+(x_1,x_2,x_3)\vert_{\rm 2\,loop,\,loc}^{(3)}
= - \frac{3i\mu^{-2\e} \beta \lambda^2 \Gamma^2(1-\e/2)}{ 2^8 \pi^{4-\e} (1-\e)^2} a_1^d \left(-\frac{1}{\e^2}+\ln^2 a_1 \right)\delta^d(x_1-x_2)\delta^d(x_1-x_3)
\label{cv7}
\end{eqnarray}

\noindent
Combining now \ref{cv5'}, \ref{cv6'} and \ref{cv7}, as well as introducing the two loop cubic vertex counterterm, 
\begin{eqnarray}
\delta \beta_{\rm 2\, loop}= \delta \beta^{s}_{\rm 2\, loop}+\delta \beta^{t}_{\rm 2\, loop}+\delta \beta^{u}_{\rm 2\, loop}=  \frac{3\mu^{-2\e} \beta \lambda^2 \Gamma^2(1-\e/2)}{ 2^6 \pi^{4-\e} \e^2 (1-\e)^2},
\label{cv7'}
\end{eqnarray}
we finally obtain the purely local part of the renormalised cubic vertex function at two loop 
\be
-i V_+(x_1,x_2,x_3)\vert_{\rm 2\, loop,\,Ren.,\,loc.}=- \frac{3i \beta \lambda^2 }{ 2^6 \pi^{4}} a_1^4 \ln^2 a_1\delta^4(x_1-x_2)\delta^4(x_1-x_3)
\label{cv8}
\ee
where as earlier,  counterterm corresponding to each channel is equal.
\\

\noindent
Combining now \ref{cv8} with \ref{cv1} and \ref{cv3'}, we have the renormalised expression for the purely local part of the cubic vertex function
\be
-i V_+(x_1,x_2,x_3)\vert_{\rm Ren.,\,loc.}=\left(-i\beta + \frac{3i\lambda \beta}{2^4\pi^2}\ln a_1- \frac{3i \beta \lambda^2 }{ 2^6 \pi^{4}}\ln^2 a_1\,+\,{\cal O}(\beta \lambda^3) \right)a_1^4 \delta^4(x_1-x_2)\delta^4(x_1-x_3)
\label{cv9}
\ee
As of the quartic case, \ref{v15}, we identify now the effective spatially homogeneous cubic vertex function containing the secular logarithms, 
\be
\beta_{\rm eff,\,loc}(a) = \beta - \frac{3\lambda \beta}{2^4\pi^2}\ln a+ \frac{3 \beta \lambda^2 }{ 2^6 \pi^{4}}\ln^2 a\,+\,{\cal O}(\beta \lambda^3)
\label{cv10}
\ee
We wish to emphasise once again that neither \ref{v15} nor \ref{cv10} has any flat spacetime analogue ($a=1$). Also, neither of these perturbative results can be trusted after sufficiently large number of $e$-foldings, when the secular logarithm terms become large. Accordingly, we wish to attempt below obtaining non-perturbative expressions for these vertex functions, as follows.

%%%%%%%
\section{Attempting non-perturbative vertex functions from the perturbative results}\label{s5}
%%%%%%

The  secular logarithms appearing in $\lambda_{\rm eff,\,loc}(a)$ (\ref{v15}) and $\beta_{\rm eff,\,loc}(a)$ (\ref{cv10}) show breakdown of perturbation theory at late times.   In this section we wish to make an attempt, perhaps a bit heuristic, in order to find out non-perturbative values out of these results. This method will chiefly be based upon some notion of correspondence.  We shall also discuss and compare our results with the ones found from the recently proposed resummation methods inspired from, or based upon variants or  modifications of the renormalisation group (RG) suitable for the de Sitter~\cite{Miao:2021gic, Kamenshchik:2020yyn, Glavan:2021adm, Litos:2023nvj, Glavan:2023lvw, Glavan:2023tet, Miao:2024nsz}, and references therein. 

We first recall that we are dealing here with the secular logarithms arising from the local part of any diagram. By {\it local}, we essentially refer to the part of a loop containing appropriate $\delta$-functions which shrink it to a single point, as has been discussed and computed in the preceding two sections. As far as the local vertex functions are concerned, only such logarithms need to be considered. Now, such logarithms are not the leading  logarithms which the stochastic formalism  resums successfully, arising from the non-local part of a diagram~\cite{Starobinsky:1994bd}.  The power of such leading logarithms in a diagram  equals the total number of internal lines present in it and it is maximum.  While computing a correlation function, or a vacuum graph, the subleading local, and the leading logarithms appear as a simple linear combination with constant coefficients and hence naturally one  considers the leading ones only, for a resummation procedure. However, while considering an {\it amputated} diagram like the self energy or the vertex function, the leading and subleading logarithms come with a natural, qualitative distinction, by the virtue of relevant $\delta$-function(s). For example, for the self energy with quartic self interaction, such distinction can be seen in~\cite{Brunier:2004sb}. Also for our case, e.g.  \ref{v4'} shows a natural distinction between the local vertex part and the channel part at  one loop. In flat spacetime ($a=1$) the local parts are purely ultraviolet divergent constants, and hence can be renormalised away. However, for dynamical backgrounds like the de Sitter, we need to attempt some resummation scheme in order to understand the physics of such local contributions.

In order to  emphasise the physical importance of these   local logarithms of subleading powers, let us consider the Schwinger-Dyson equation, e.g.~\cite{Peskin}, in the de Sitter spacetime
\be
\left[\nabla_{\mu}\nabla^{\mu}   - m_0^2 -\ {\rm counterterms} \right]i G_{++}(x,x') = \frac{i\delta^d(x-x')}{a^d} + i\int_{x''}  (-i\Sigma_{++}(x,x'')) i G_{++}(x'',x) 
\label{addR1}
\ee
where $i G_{++}(x,x') $ is the exact Feynman propagator and $-i\Sigma_{++}(x,x'')$ is the sum of 1PI self energies up to some perturbative order. After renormalisation, the local part of the self energy is given by $-i\Sigma_{++}^{\rm loc.}(x,x'') = f(\ln a ) \delta^d(x-x'') $
(say), so that in the renormalised equation, the mass term is shifted as 
$$m_0^2 \to m_0^2 + f(\ln a)$$
where $f(\ln a)$ is a function of the local secular logarithms. Since the rest mass of the field is essentially a localised quantity, it is natural to identify the term   $f(\ln a)$ as the dynamically generated rest mass of the field. Since at late times $\ln a$ grows unboundedly, we need to employ  some resummation to extract a physically meaningful result.  This can be achieved by, for example, by promoting the propagators appearing in the self energy diagrams by the exact propagators via infinite number of self energy insertions, e.g.~\cite{Youssef:2013by} and references therein. We assume that the corresponding resummed result is finite.

Let us consider the self energy of $\lambda \phi^4$ theory now~\cite{Brunier:2004sb}. The ${\cal O}(\lambda)$ contribution is purely local, proportional to  $\sim \lambda i\Delta_{++}(x,x)$, which is $\sim\lambda \ln a $. Note from \ref{y6} that even though local, this secular logarithm is {\it not} associated
with the ultraviolet divergent term. On the other hand at ${\cal O} (\lambda^2)$ there are two contributions -- the snowman and the sunset. For the first, we have an integral like 
$$\sim \int a'^d d^d x' i\Delta^2_{++}(x,x') i\Delta(x',x')$$
Thus there will be local contributions from  $i\Delta_{++}(x,x)$, as well as from \ref{e51'} due to the square of the propagator. Likewise for the sunset diagram, the local part of the self energy comes from the cube of the propagator, altogether making a $\sim \lambda^2 \ln^2 a$ contribution to the dynamical mass in \ref{addR1}. The local logarithms arising from products of propagators (such as \ref{e51'}) essentially originate from the volume measure of the spacetime integral and the van Vleck-like determinant $(aa')^{-1+\e/2}$ coming from the denominator of the $A(x,x')$ term of \ref{props2}. Thus such logarithms may perhaps be interpreted as the manifestation  of the spacetime expansion, and not any artifact of the dimensional regularisation procedure\footnote{Note that such logarithms associated with the determinant of the metric may also arise in the standard Schwinger-DeWitt type expansion  of the propagator $iG_{++}(x,x')$, with respect to the Riemann normal coordinatisation around $x$~\cite{Parker:2009uva}. However, for any local contribution the $\delta$-function makes $x'$  merge with $x$, where $\sqrt{-g}$ equals unity.}.    Nevertheless, all such local logarithms, no matter what there origin is, contribute in an equal footing to the dynamical mass of the field. This suggests that they must be resummed in an equal footing. However, we cannot use any stochastic method to capture any purely local effect and instead need to apply the Hartree or Hartree-like scheme of quantum field theory.  

We shall come to the issue of explicit resummation, but let us first note a connection between the local self energy and the correction to the coupling constant. We modify \ref{addR1} with a background field, say $v$, along with the quantum field $\phi$. First, this will generate an additional term, $-\lambda v^2/2$ on the left hand side within the square bracket. Second, there will now be a cubic self interaction term, $\lambda v \phi^3/3!$. The one loop self energy  corresponding to this self interaction reads
\be
-\frac{\lambda^2 v^2(aa')^d}{2} i\Delta_{++}^2(x,x') 
\label{addR2}
\ee
Using \ref{e51'} and after renormalisation using a vertex counterterm, we have the replacement by the virtue of the local part
$$ \frac{\lambda v^2}{2} \to \frac{v^2}{2}\left(\lambda - \frac{\lambda^2 \ln a}{2^3 \pi^2} \right) $$
in \ref{addR1}. Comparing this with the one loop part of \ref{v15}, we see mismatch in numerical factors. This is not surprising, as we have not computed any amputated vertex diagram here, although we may note the qualitative similarity between these two expressions, for both being originated from the square of the propagator. The above relationship shows that the renormalised local self energy actually modifies the coupling. As the resummed local self energy is related to the dynamically generated field rest mass, we may  attempt to obtain non-perturbative values of the vertex functions computed in this paper, via the dynamical mass. We also note that if the field developes a rest mass at late times, it should be reflected in the vertex functions as well.

Even though we have restricted ourselves to quartic self interaction above, we may easily include the cubic self interaction as well. For example, considering the one loop self energy at ${\cal O}(\beta \lambda v)$ shows how the renormalised cubic coupling gets modified. 

Let us now come to the explicit issue of resummation of the secular local logarithms. At ${\cal O}(\lambda)$, we have a term $\sim \lambda \ln a  $, which is proportional to $\lambda \langle \phi^2 \rangle_{\rm Ren.} $,  \ref{y6}. We promote this to non-perturbative level, by replacing the propagator of the vacuum bubble by the exact one. Let the dynamically generated mass at late times be $m^2_{\rm dyn.}$. Assuming $m^2_{\rm dyn.}/H^2 \ll 1$, we have the well known result for a free light scalar~\cite{Allen:1985ux}  
\be
i\Delta_{++}(x,x,m^2_{\rm dyn.})\vert_{\rm Ren.} = \langle \phi^2 \rangle_{\rm Ren.} = \frac{3H^4}{8\pi^2 m^2_{\rm dyn.}}
\label{rs3}
\ee
Equating now the one loop local self energy $\lambda\langle \phi^2 \rangle_{\rm Ren.}/2 $ to $m^2_{\rm dyn.}$ in \ref{addR1}, we have from the above expression
\be
{m}^2_{\rm dyn.}= \frac{\sqrt{3\lambda} H^2}{4\pi} 
\label{rs3'} 
\ee 
which is the same as we obtain using the Hartree or local approximation. This approximation does {\it not} resum the leading non-local logarithms. The value of $\langle \phi^2\rangle$  appearing in  \ref{rs3} is purely local and does not equal the stochastically resummed one.  

Equivalently, we consider the correlation function  $\langle \phi^2 \rangle$ for a massless minimal scalar up to ${\cal O}(\lambda)$,
\be
\langle \phi^2 \rangle_{\rm Ren.} = \frac{H^2 \ln a}{4\pi^2} - \frac{\lambda \ln^3 a}{2^4\times 9\pi^4} +{\cal O}(\lambda^2)
\label{addR5}
\ee
We differentiate both sides with respect to $\ln a$ and next replace the $\ln^2 a $ appearing on the right hand side by $\sim \langle \phi^2\rangle^2$. At late times we expect to reach an equilibrium when $\langle \phi^2\rangle$ is given by \ref{rs3}, so that the left hand side of this differential equation is vanishing then and we reproduce \ref{rs3'}.

In the presence of a cubic coupling, and going up to ${\cal O}(\lambda \beta^2)$ of the self energy, one can derive the algebraic equation~\cite{Bhattacharya:2022wjl, Bhattacharya:2023xvd},  
\be 
(\bar{m}_{\rm dyn}^2)^3 = \frac{6\lambda - 3\bar{\beta}^2}{32\pi^2} \bar{m}_{\rm dyn}^2 -\frac{9\lambda(\lambda -2 \bar{\beta}^2)}{256 \pi^4} \qquad (\bar{m}_{\rm dyn}^2 \geq 0)
\label{rs5}
\ee
where the bar over the quantities denote scaling with respect to the Hubble rate $H$.  Setting $\bar{\beta}=0$ above  reproduces \ref{rs3'} at the leading order. On the other hand for $\lambda =0$, we have $\bar{m}_{\rm dyn}^2=0$.  This is consistent with the rolling down and runaway disaster expected in a pure cubic potential.  If we now `turn on' a small $\lambda$ value, we have from \ref{rs5}, 
\be
\bar{m}_{\rm dyn}^2= \frac{3\lambda}{4\pi^2}\left(1+ {\cal O}(\lambda/\bar{\beta}^2) \right)   
\label{rs5''}
\ee
The above result  is a manifestation of the fact that the  quartic (with $\lambda>0$) {\it plus} cubic self interaction is always bounded from below irrespective of the value of $\lambda$, and hence there can be no runaway disaster as of the pure cubic case.

To  summarise, the above results are based upon two key things. The first is considering only the local logarithms and treat them in an equal footing, as dictated by \ref{addR1}. Second, we have replaced each $\ln a$ by the vacuum bubble, and have made them non-perturbative via \ref{rs3}, which is a purely local and non-stochastic relationship.    The promotion from perturbative to non-perturbative propagators  can explicitly be thought of as equivalent to generating daisy-like Feynman graphs, by infinite number of self energy insertions, e.g.~\cite{Burgess:2015ajz, Youssef:2013by}.            \\

\begin{figure}[htbp]
    \centering
    \includegraphics[width=17cm, height=5cm]{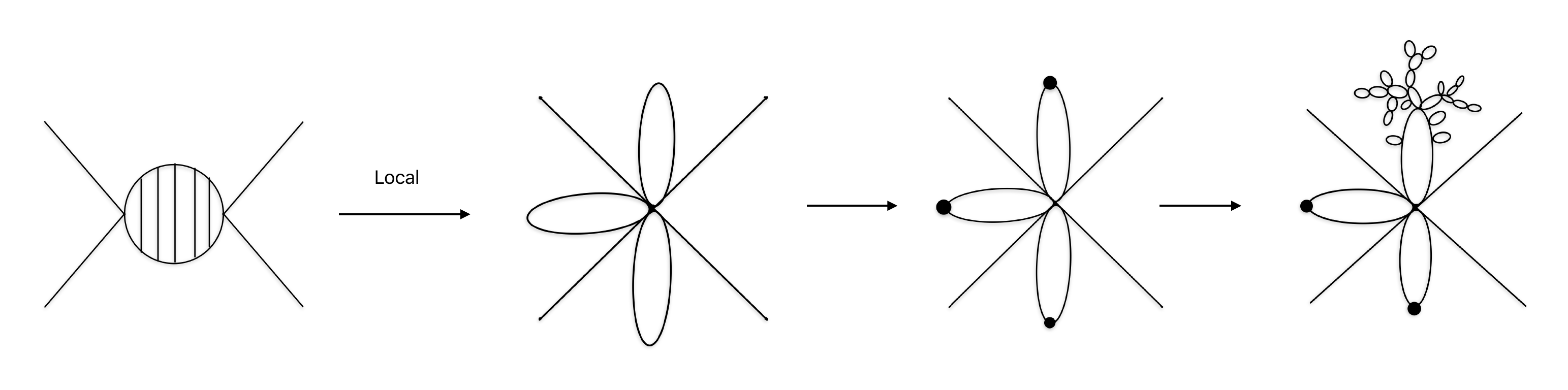}
    \caption{\small \it The resummed quartic vertex function via the dynamical mass generation from purely local contributions. The second diagram correspond to the local part of the vertex function containing power of the secular local logarithm, $\ln a $, with all the vertex points coincident. Each such logarithm is replaced in the favour of $\langle \phi^2 \rangle$. $\langle \phi^2 \rangle$ is then promoted to non-perturbative level.  A dot over a propagator indicates that it is exact. One possible way to realise this perturbative to non-perturbative promotion is to represent an exact propagator by a daisy-like Feynman graph, containing infinite number of loop insertions in the free propagator, then taking the local parts of those loop contributions and once again promoting them to the non-perturbative level, and so on. Note in particular that we do not use any stochastically resummed value of $\langle \phi^2 \rangle$ here, which contains contributions from  non-local self energy. Instead we use here \ref{rs3} for the non-perturbative bubble. This is purely local and not stochastic.  Similar arguments hold for the resummed cubic vertex as well. See main text for discussion. }
    \label{daisy}
\end{figure}

\begin{figure}[htbp]
    \centering
    \begin{subfigure}[b]{0.42\textwidth}
        \centering
        \includegraphics[width=\textwidth]{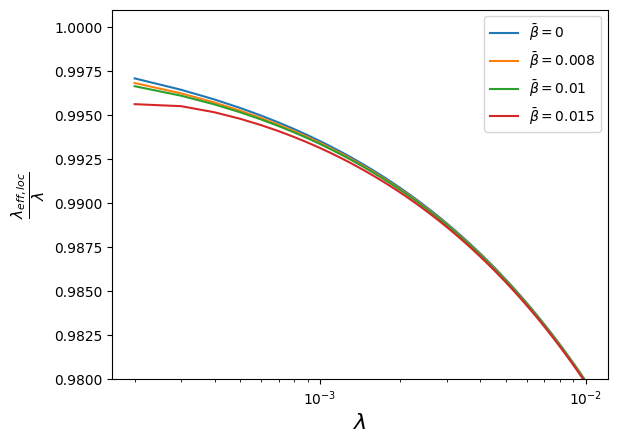}
        %\caption{Figure 1}
    \end{subfigure}
    \hspace{2cm}
    \begin{subfigure}[b]{0.42\textwidth}
        \centering
        \includegraphics[width=\textwidth]{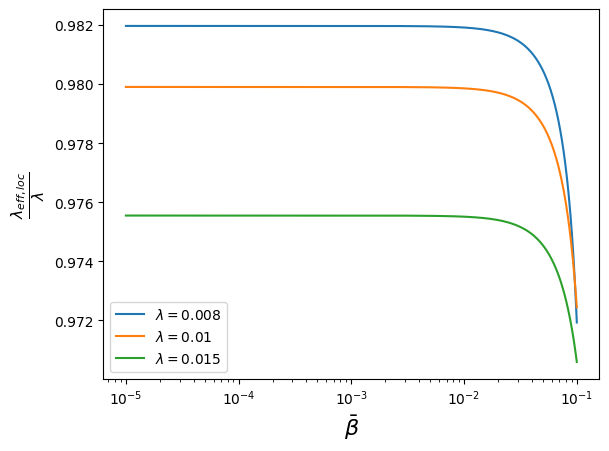}
        %\caption{Figure 2}
    \end{subfigure}
    \caption{\small \it Variation of the resummed value of the quartic vertex function, \ref{rs4}, with respect to the tree level coupling parameters. See main text for discussion. }
    \label{4-plot}
\end{figure}

\noindent
Let us now come to the chief concern of this paper -- the local vertex functions. We already  have argued above that we may associate with them the local self energy and hence the dynamical mass.   Accordingly,  we replace each $\ln a$ appearing in   \ref{v15} by $\sim \langle \phi^2\rangle$ and promote it to non-perturbative level by the virtue of \ref{rs3},    to yield
\be 
\frac{\lambda_{\rm eff,\,loc}}{\lambda}= 1 -\frac{9\lambda}{32 \pi^2 \bar{m}^2_{\rm dyn}}  + \frac{135\lambda^2 }{1024 \pi^4 \bar{m}^4_{\rm dyn}  } 
\label{rs4}
\ee
where $\bar{m}^2_{\rm dyn}$ is given by \ref{rs5}. We have attempted to depict   the whole procedure schematically  in \ref{daisy}.  
Now for the quartic self interaction only ($\beta=0$),  we have from \ref{rs3'}
\be 
\frac{\lambda_{\rm eff,\,loc}}{\lambda}\Big\vert_{\bar{\beta}\to 0}= 1 -\frac{3\sqrt{3\lambda}}{8\pi}  + \frac{45\lambda }{64 \pi^2} +{\cal O}(\lambda^{3/2})
\label{rs6}
\ee
Clearly, the above series is converging for $\lambda < 1$. Thus if we take for example, $\lambda \sim 0.1$, we have from \ref{rs6}
$$\frac{\lambda_{\rm eff,\,loc}}{\lambda}\Big\vert_{\bar{\beta}\to 0}\approx 1- 0.058 $$
indicating about  $6\%$ decrease in the value to the quartic coupling at late times. For any $\lambda \lesssim {\cal O}(1)$, there will be reduction in the quartic coupling. On the other hand for $\lambda \to 0$, we have from \ref{rs4}, \ref{rs5''}
\be 
\frac{\lambda_{\rm eff,\,loc}}{\lambda}\Big\vert_{\lambda \to 0}\approx  1 -0.083,
\label{rs6'}
\ee
showing around  $8\%$ decrease, irrespective of the value of any coupling, as dictated by \ref{rs5''}.  We have depicted the variation of $\lambda_{\rm eff,\,loc}/\lambda$, \ref{rs4}, in \ref{4-plot} with respect to the tree level coupling parameters. 
\\

\noindent 
The resummation of the vertex function can also be performed using different methods chiefly related to RG. We now wish to compare the above result with that of found via such methods. \\

\noindent
We first consider a resummation technique inspired by the renormalisation group, proposed recently in~\cite{Kamenshchik:2020yyn, Kamenshchik:2021tjh}. This formalism was applied recently in~\cite{Bhattacharya:2022aqi, Bhattacharya:2022wjl}, in order to compute the non-perturbative vacuum expectation values of the scalar field $\phi$ as well the dynamically generated mass. We rewrite \ref{v15} as
\be 
\Delta \lambda= \frac{3\lambda^2 {\cal N}}{2^4 \pi^{2}} - \frac{15\lambda^3{\cal N}^2 }{2^8\pi^{4} } 
\label{rs7}
\ee
where ${\cal N}=Ht$ is the number of de Sitter $e$-foldings and  we have abbreviated
$$\Delta \lambda = -\left(\lambda_{\rm eff,loc}(a) -\lambda\right)$$
Following~\cite{Kamenshchik:2020yyn, Kamenshchik:2021tjh}, we differentiate \ref{rs7} once with respect to ${\cal N}$, 
and next using \ref{rs7}, we replace the $e$-folding in the favour of $\Delta \lambda$, so that we have  
\be 
\frac{\partial \Delta \lambda}{\partial {\cal N}}= -\frac{5\lambda}{2^3 \pi^{2}} \left(\Delta \lambda -\frac{3 \lambda}{10} \right)
\label{rs9}
\ee
We integrate the above equation and find as ${\cal N}\gg 1 $, we must have 
\be 
\lambda_{\rm eff,\,loc}= \lambda -0.3\lambda
\label{rs10}
\ee
which shows $30\%$ decrease in the quartic coupling, which seems to be rather high.  Analogous result holds    for the cubic vertex function, \ref{cv10}, as well. We are not sure whether \ref{rs10} is a physically acceptable result. To the best of our understanding,  it  seems that the resummation proposal  of~\cite{Kamenshchik:2020yyn, Kamenshchik:2021tjh} will give satisfactory results if we attempt to deal with the expectation value of some operator, which satisfies some equation of motion, such as $\langle \phi \rangle$ and $\langle \phi^2 \rangle$, as of~\cite{Bhattacharya:2022aqi, Bhattacharya:2022wjl}. In other words, the same resummation procedure is likely to work well when we compute any correlation function of the scalar, $\langle \phi^n \rangle $.  \\

\noindent
Finally, we come to the RG method to address the resummation of de Sitter secular logarithms, developed in a series of works very recently~\cite{Miao:2021gic, Glavan:2021adm, Litos:2023nvj, Glavan:2023lvw, Glavan:2023tet, Miao:2024nsz}. This is inspired from the renormalisation scale dependence of ultraviolet finite quantities, such as the last term appearing on the right hand side of the second line of \ref{e51'}.  However, there exists a fundamental difference between the RG in the flat spacetime (e.g.~\cite{Peskin}) with  that of the de Sitter that, for the latter the renormalisation scale $\mu$ is continually changing owing to the rapid expansion of the spacetime. Accordingly, a novel idea was introduced in these works to replace the logarithm, $\ln \mu$ appearing in the Callan-Symanzik equation or the $\beta$-functions by $\ln (\mu a)$ or $\ln (\mu a/H)$, should we wish to keep the argument  dimensionless. Using this modification, various resummation of the secular logarithms was performed recently in~\cite{Miao:2024nsz}, including that of the quantum corrected gravitational potential and the Weyl tensor and those which cannot be captured by the stochastic method. We wish to use this de Sitter-modified RG technique below to resum the logarithms of the local quartic vertex function. We shall restrict our analysis to one loop order only.

Thus from  \ref{v3}, we have the one loop $\beta$-function
\be
\beta_{\lambda} = \frac{\p \lambda_{\rm eff,loc. }}{\p \ln (\mu/Ha)} =\frac{3\lambda^2}{16\pi^2}
\label{addR2'}
\ee
The above equation yields 
\be
 \lambda_{\rm eff,loc.} (Ha/\mu) = \frac{\lambda}{1  + \frac{3\lambda}{16\pi^2}  \ln \frac{Ha}{\mu}  }
\label{addR4}
\ee
Note that the term $Ha$ can be interpreted as the  scale of a super-Hubble comoving momentum at any given instance.  Note also that whether $\lambda_{\rm eff}$'s value would be less or greater than $\lambda$ will depend upon the ratio $Ha/\mu$. After large $e$-foldings, we may take $Ha>\mu$, and hence $\lambda_{\rm eff, loc.} < \lambda$ then.

Comparing \ref{addR4} with \ref{rs4}, we note the absence of the renormalisation scale $\mu$ there. This corresponds to our renormalisation scheme that the scale should be kept in the counterterms. We could have alternatively kept the scale in the renormalised quantities, by modifying the counterterms. As we have mentioned above, this can be motivated from the $\mu$-dependent ultraviolet finite term appearing in \ref{e51'}. Also, the denominator of \ref{addR4} shows an infinite binomial series, probably corresponding to the series of diagrams at one loop, only the first diagram of   \ref{4-2l}, a similar diagram at three loop and so on. Finally, and most importantly, the existence of the secular logarithm shows $\lambda_{\rm eff, loc.}$ to be vanishing at sufficiently late times. This is analogous to the flat spacetime result with $Ha$ replaced by the momentum. Can we replace it via the dynamically generated mass of the field at late times in \ref{addR4}? To the best of our understanding, and as we have tried to argue, Yes. This is because unlike the flat spacetime, the theory generates a scale in de Sitter in terms of the dynamical mass.  Alternatively, \ref{addR4} suggests that the $\phi^4$-theory becomes asymptotically free at sufficiently late times in de Sitter.   \\

\begin{figure}[htbp]
    \centering
    \begin{subfigure}[b]{0.42\textwidth}
        \centering
        \includegraphics[width=\textwidth]{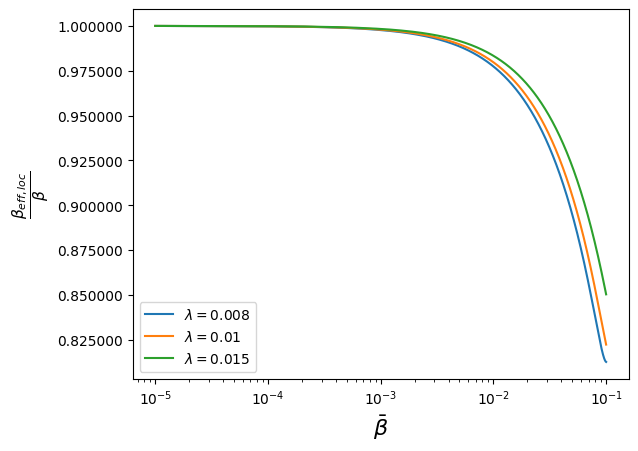}
    \end{subfigure}
    \hspace{2cm}
    \begin{subfigure}[b]{0.42\textwidth}
        \centering
        \includegraphics[width=\textwidth]{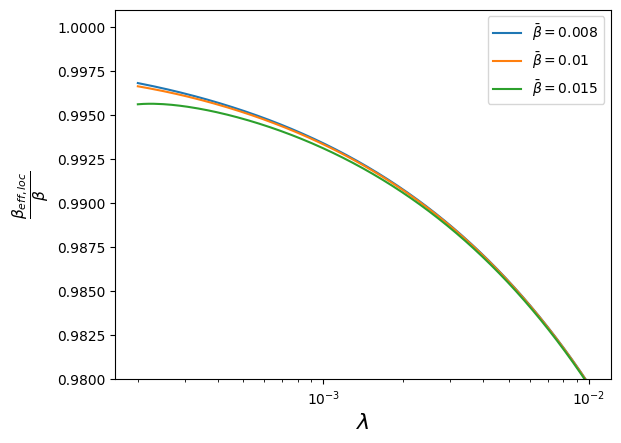}
    \end{subfigure}
    \caption{\small \it Variation of the resummed value of the cubic vertex function, \ref{rs6}, with respect to the tree level coupling parameters. See main text for discussion.  }
    \label{3-plot}
\end{figure}

\noindent
For the cubic vertex function, \ref{cv10}, we obtain as per our resummation scheme associated with the dynamical mass  
\be 
\frac{\bar{\beta}_{\rm eff,\,loc}}{\bar{\beta}}= 1 -\frac{9\lambda}{32 \pi^2 \bar{m}^2_{\rm dyn}}  + \frac{27\lambda^2 }{256 \pi^4 \bar{m}^4_{\rm dyn}} 
\label{rs6''}
\ee
showing both the effective quartic and cubic coupling strengths  have similar behaviour with respect to the variation of the tree level coupling parameters, $\lambda$ and $\bar{\beta}$, as depicted in \ref{3-plot}.\\

%

%%%%%%
\section{Discussion}\label{s6}
%%%%%% 
In this paper we have considered the effect of the late time non-perturbative  secular growth present in a self interacting, massless and minimally coupled quantum scalar field on the vertex functions in the inflationary de Sitter background. We have considered quartic and quartic plus cubic self interaction potentials in four spacetime dimensions, and have considered 4- and 3-point vertices. We have done explicit computations up to two loop for these vertex functions and have focused on the purely local part of them, in which case the vertices shrink to single points.  The corresponding perturbative expressions for the vertex functions are given respectively by \ref{v15}, \ref{cv10}.  These local logarithms have sub-leading powers compared to the leading, non-local ones and cannot be resummed via the stochastic method.  We have next argued and assigned resummed values to the secular logarithms of these vertex functions in terms of the dynamically generated late time mass of the scalar field in \ref{s5}.  Since the dynamical mass essentially involves non-perturbative resummation of the local part of the self energy, the results we have obtained for the vertex functions are also non-perturbative, as we have argued in \ref{s5} (\ref{rs4}, \ref{rs6''}). As we have also mentioned in the main text of this paper, certainly this phenomenon can have no flat spacetime analogue.  We also have computed briefly the resummed quartic vertex function via a method inspired from RG~\cite{Kamenshchik:2020yyn, Kamenshchik:2021tjh}, or an RG method suitable for de Sitter~\cite{Miao:2021gic, Glavan:2021adm, Litos:2023nvj, Glavan:2023lvw, Glavan:2023tet, Miao:2024nsz}, and have compared various results.  All these results in particular, suggest that at late times the value of the non-perturbative vertex function should be less than the tree level coupling.   
      
What happens to the partly local and purely local vertex functions, appearing in, for example in \ref{v6}?  At least for the partly local contributions (i.e., those with channel structure and without any integration), there are logarithms of the scale factor. Such logarithms appear from the Taylor expansion of a term like $a^{d\pm \e}/\e$. On the other hand, such term also originates from the local part of the square of the Feynman propagator, \ref{e51'}. Thus it seems that we may replace these logarithms by the dynamical mass as well. Nevertheless, the full computation of any such vertex function which is not  purely local must involve the in-in or the Schwinger-Keldysh formalism owing to the non-coincidence expressions for the propagator. In this case we will need all the four kind of propagators mentioned in \ref{s2}~\cite{Calzetta, Hu, Adshead}. Apart from this, it seems also an interesting task to understand the results found in this paper from the perspective of a non-perturbative effective action. We wish to come back to this issue in a future work.

%%%%
\section*{Acknowledgement} The authors would like to sincerely acknowledge anonymous referee for a careful, critical reading of the manuscript and for making various valuable comments and suggestions, in particular regarding the resummation procedure, on an earlier version of this manuscript. 
%%%%

\bigskip
\bigskip
\appendix
\labelformat{section}{Appendix #1} 

%%%%%%%%%%
\section{The diagrams with purely non-local contributions but no non-vanishing flat spacetime limit}\label{A}
\begin{figure}[htbp]
    \centering
    \includegraphics[width=15cm, height=5cm]{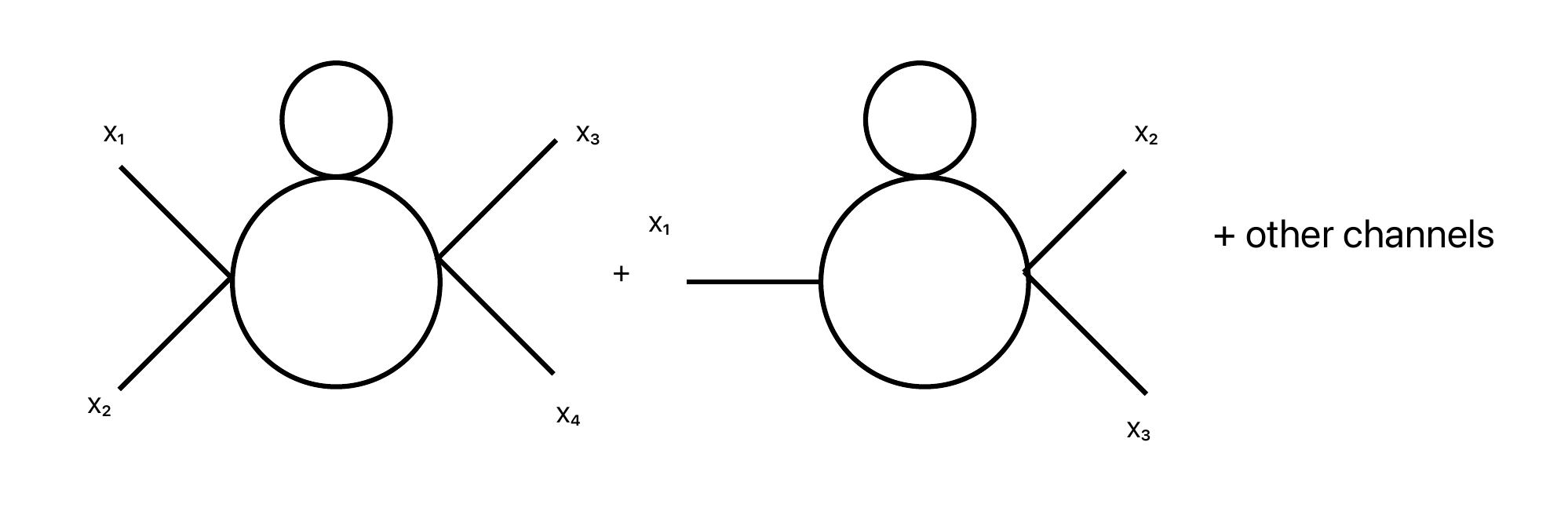}
    \caption{\small \it Two loop diagrams which are vanishing after renormalisation in flat spacetime, but non-vanishing in de Sitter due to the secular logarithms. However, these diagrams do not yield any purely local contribution to the quartic or cubic vertex functions and hence do not affect the main results of this paper. }
    \label{fA1}
\end{figure}

\noindent
In this appendix, we wish to discuss the diagrams, \ref{fA1}, which do not yield any purely local contributions to the vertex functions, yet they yield some non-local contribution by the virtue of the secular logarithm.  For the quartic vertex, the corresponding contribution at two loop reads
\begin{eqnarray}
&&-i V_+(x_1,x_2,x_3, x_4)=  \frac{i\lambda^3}{2^2} (a_1a_2)^d \int d^d x a^d\, i\Delta (x,x) i\Delta_{++}(x_1,x)i\Delta_{++}(x,x_2) i\Delta _{++}(x_1,x_2) \delta^d(x_1-x_3)\delta^d (x_2-x_4) \nonumber\\&& +
\frac{i\lambda^3}{2^2} (a_2 a_3)^d \int d^d x a^d \,i\Delta (x,x) i\Delta_{++}(x_2,x)i\Delta_{++}(x,x_3) i\Delta _{++}(x_2,x_3) \delta^d(x_1-x_2)\delta^d (x_3-x_4) \nonumber\\&&
  +\frac{i\lambda^3}{2^2} (a_3a_1)^d \int d^d x a^d\, i\Delta (x,x) i\Delta_{++}(x_1,x)i\Delta_{++}(x,x_3) i\Delta _{++}(x_1,x_3) \delta^d(x_1-x_4)\delta^d (x_2-x_3) 
\label{A1}
\end{eqnarray}

We add to the above the one loop mass renormalisation counterterm (due to the quartic self interaction) diagram contribution
\begin{eqnarray}
&&-i V_+(x_1,x_2,x_3, x_4)\vert_{\rm CT}=  \frac{i\lambda^2 \delta m_{\lambda}^2}{2} (a_1a_2)^d \int d^d x a^d\, i\Delta_{++}(x_1,x)i\Delta_{++}(x,x_2) i\Delta _{++}(x_1,x_2) \delta^d(x_1-x_3)\delta^d (x_2-x_4) \nonumber\\&& +
\frac{i\lambda^2 \delta m_{\lambda}^2}{2} (a_2 a_3)^d \int d^d x a^d \, i\Delta_{++}(x_2,x)i\Delta_{++}(x,x_3) i\Delta _{++}(x_2,x_3) \delta^d(x_1-x_2)\delta^d (x_3-x_4)  \nonumber\\&&
  +\frac{i\lambda^2 \delta m_{\lambda}^2}{2} (a_3a_1)^d \int d^d x a^d\,  i\Delta_{++}(x_1,x)i\Delta_{++}(x,x_3) i\Delta _{++}(x_1,x_3) \delta^d(x_1-x_4)\delta^d (x_2-x_3)
  \label{A2}
\end{eqnarray}
Substituting now \ref{y6} for $i\Delta(x,x)$ into \ref{A1}, and \ref{y7} into \ref{A2} for $\delta m_{\lambda}^2$, we see that the the divergence of \ref{A1} cancels, leaving us with an ultraviolet finite integral 
\begin{eqnarray}
&&-i V_+(x_1,x_2,x_3, x_4)\vert_{\rm 2\,loop,\,Ren.}^{(4)}=  \frac{i\lambda^3 H^2}{2^4\pi^2} (a_1a_2)^d \int d^4 x a^4\, \ln a\, i\Delta_{++}(x_1,x)i\Delta_{++}(x,x_2) i\Delta _{++}(x_1,x_2) \delta^d(x_1-x_3)\delta^d (x_2-x_4) \nonumber\\&& +
\frac{i\lambda^3 H^2}{2^4 \pi^2} (a_2 a_3)^d \int d^4 x a^4 \, \ln a \,i\Delta_{++}(x_2,x)i\Delta_{++}(x,x_3) i\Delta _{++}(x_2,x_3) \delta^d(x_1-x_2)\delta^d (x_3-x_4) \nonumber\\&&
  +\frac{i\lambda^3 H^2}{2^4\pi^2} (a_3a_1)^d \int d^4 x a^4\, \ln a \,i\Delta_{++}(x_1,x)i\Delta_{++}(x,x_3) i\Delta _{++}(x_1,x_3) \delta^d(x_1-x_4)\delta^d (x_2-x_3) 
\label{A3}
\end{eqnarray}

It is easy to see that the above integral does not yield any purely local contribution. Also, the above integral vanishes in the flat spacetime limit, $a=1$.  \\

\noindent
Likewise for the cubic vertex, i.e. the second of \ref{fA1}, we have 
\begin{eqnarray}
&&-i V_+(x_1,x_2,x_3)\vert_{\rm 2\,loop}=  \frac{i\beta \lambda^2}{2^2} (a_1a_2)^d \int d^d x a^d\, i\Delta (x,x) i\Delta_{++}(x_1,x)i\Delta_{++}(x,x_2) i\Delta _{++}(x_1,x_2) \delta^d(x_2-x_3) \nonumber\\&& +
\frac{i\lambda^3}{2^2} (a_2 a_3)^d \int d^d x a^d \,i\Delta (x,x) i\Delta_{++}(x_2,x)i\Delta_{++}(x,x_3) i\Delta _{++}(x_2,x_3) \delta^d(x_1-x_3) \nonumber\\&&
  +\frac{i\lambda^3}{2^2} (a_3a_1)^d \int d^d x a^d\, i\Delta (x,x) i\Delta_{++}(x_1,x)i\Delta_{++}(x,x_3) i\Delta _{++}(x_1,x_3) \delta^d(x_1-x_2)
\label{A4}
\end{eqnarray}
which can be renormalised as of \ref{A1} to yield an integral which has only non-local contribution, via the secular logarithm. Both the above and \ref{A3} shows a qualitative difference of quantum field theory result of the de Sitter and the flat spacetime.  

\bigskip
%%%%%%%%%%%%%%%%%%%%%%%%%%%%%%%%


\begin{thebibliography}{99} 


\bibitem{Wein}
S. Weinberg, 
{\it Cosmology}, Oxford Univ. Press (2009).
%
\bibitem{Mukhanov:2005sc}
V.~Mukhanov,
{\it Physical Foundations of Cosmology},
Cambridge University Press, 2005.
%


%
\bibitem{Tsamis}
N. C. Tsamis and R. P. Woodard, 
{\it Relaxing the cosmological constant}, Phys. Lett. B301, 351 (1993).
%
\bibitem{Ringeval}
C. Ringeval, T. Suyama, T. Takahashi, M. Yamaguchi and S. Yokoyama, 
{\it Dark energy from primordial infationary quantum 
fluctuations},
Phys. Rev. Lett.105, 121301 (2010) [arXiv:1006.0368 [astro-ph.CO]].
%

%
\bibitem{Miao:2021gic}
S.~P.~Miao, N.~C.~Tsamis and R.~P.~Woodard,
{\it Summing inflationary logarithms in nonlinear sigma models},
JHEP \textbf{03}, 069 (2022)
[arXiv:2110.08715 [gr-qc]].
%


\bibitem{Dadhich}
N. Dadhich, 
{\it On the measure of spacetime and gravity},
Int. J. Mod. Phys. D20, 2739-2747 (2011)
[arXiv:1105.3396 [gr-qc]].
%
\bibitem{Padmanabhan}
T. Padmanabhan and H. Padmanabhan, CosMIn: {\it The Solution to the Cosmological Constant Problem},
Int. J. Mod. Phys. D22, 1342001 (2013) [arXiv:1302.3226 [astro-ph.CO]].
%
\bibitem{Alberte}
L. Alberte, P. Creminelli, A. Khmelnitsky, D. Pirtskhalava and E. Trincherini, 
{\it Relaxing the Cosmological
Constant: a Proof of Concept}, JHEP12, 022 (2016) [arXiv:1608.05715 [hep-th]].
%
\bibitem{Appleby}
S. Appleby and E. V. Linder, 
{\it The Well-Tempered Cosmological Constant}, JCAP07, 034 (2018) [arXiv:1805.00470 [gr-qc]].
%
\bibitem{Khan:2022bxs}
A.~Khan and A.~Taylor,
{\it A minimal self-tuning model to solve the cosmological constant problem},
JCAP \textbf{10}, 075 (2022)
[arXiv:2201.09016 [astro-ph.CO]].
%
%
\bibitem{Evnin:2018zeo}
O.~Evnin and K.~Nguyen,
{\it Graceful exit for the cosmological constant damping scenario},
Phys. Rev. D \textbf{98}, no.12, 124031 (2018)
doi:10.1103/PhysRevD.98.124031
[arXiv:1810.12336 [gr-qc]].
%
%
\bibitem{Floratos}
E. G. Floratos, J. Iliopoulos and T. N. Tomaras, 
{\it Tree Level Scattering Amplitudes in De Sitter Space diverge}, Phys. Lett. B197, 373 (1987).
%
%
\bibitem{Chernikov:1968zm}
N.~A.~Chernikov and E.~A.~Tagirov,
{\it Quantum theory of scalar fields in de Sitter space-time},
Ann. Inst. H. Poincare Phys. Theor. A \textbf{9}, 109 (1968)
%
\bibitem{Bunch:1978yq}
T.~S.~Bunch and P.~C.~W.~Davies,
{\it Quantum Field Theory in de Sitter Space: Renormalization by Point Splitting},
Proc. Roy. Soc. Lond. A \textbf{360}, 117-134 (1978).
%
\bibitem{Linde:1982uu}
A.~D.~Linde,
{\it Scalar Field Fluctuations in Expanding Universe and the New Inflationary Universe Scenario},
Phys. Lett. B \textbf{116}, 335-339 (1982).
%
\bibitem{Starobinsky:1982ee}
A.~A.~Starobinsky,
{\it Dynamics of Phase Transition in the New Inflationary Universe Scenario and Generation of Perturbations},
Phys. Lett. B \textbf{117}, 175-178 (1982).
%
\bibitem{Allen:1985ux}
B.~Allen,
{\it Vacuum States in de Sitter Space},
Phys. Rev. D \textbf{32}, 3136 (1985).
%
\bibitem{Allen}
B. Allen and A. Folacci,
{\it Massless minimally coupled scalar field in de Sitter space},
Phys. Rev. D \textbf{35}, 3771 (1987).
%
\bibitem{Karakaya:2017evp}
G.~Karakaya and V.~K.~Onemli,
{\it Quantum effects of mass on scalar field correlations, power spectrum, and fluctuations during inflation},
Phys.~Rev.~D\textbf{97}, no.12, 123531 (2018)
[arXiv:1710.06768 [gr-qc]].
%


%%%%%%%%%%%%%%%%%%%%%%%%%%%%%%%%%%%%
%for one loop two loop computation
%¯
\bibitem{Onemli:2002hr}
V.~K.~Onemli and R.~P.~Woodard,
{\it Superacceleration from massless, minimally coupled $\phi^4$},
Class.~Quant.~Grav.\textbf{19}, 4607 (2002)
[arXiv:gr-qc/0204065 [gr-qc]].
%
\bibitem{Brunier:2004sb} 
  T.~Brunier, V.~K.~Onemli and R.~P.~Woodard,
{\it Two loop scalar self-mass during inflation}.
  Class.\ Quant.\ Grav.\  {\bf 22}, 59 (2005)
  [gr-qc/0408080].
%
\bibitem{Kahya:2009sz}
E.~O.~Kahya, V.~K.~Onemli and R.~P.~Woodard,
{\it A Completely Regular Quantum Stress Tensor with $w \ensuremath{<} -1$},
Phys. Rev. D\textbf{81}, 023508 (2010)
[arXiv:0904.4811 [gr-qc]].
%
\bibitem{Boyanovsky:2012qs}
D.~Boyanovsky,
{\it Condensates and quasiparticles in inflationary cosmology: mass generation and decay widths},
Phys. Rev. D\textbf{85}, 123525 (2012)
[arXiv:1203.3903 [hep-ph]].
%
\bibitem{Onemli:2015pma}
V.~K.~Onemli,
{\it Vacuum Fluctuations of a Scalar Field during Inflation: Quantum versus Stochastic Analysis},
Phys.~Rev.~D\textbf{91}, 103537 (2015)
[arXiv:1501.05852 [gr-qc]].
%

 %
\bibitem{Prokopec:2003tm} 
 T.~Prokopec and E.~Puchwein,
{\it Photon mass generation during inflation: de Sitter invariant case},
  JCAP {\bf 0404}, 007 (2004)
  [astro-ph/0312274].
%
\bibitem{Miao:2006pn}
  S.~P.~Miao and R.~P.~Woodard,
{\it Leading log solution for inflationary Yukawa},
  Phys.\ Rev.\ D {\bf 74}, 044019 (2006)
  [gr-qc/0602110].
 % 
 \bibitem{Prokopec:2007ak}
T.~Prokopec, N.~C.~Tsamis and R.~P.~Woodard,
{\it Stochastic Inflationary Scalar Electrodynamics},
Annals Phys.\textbf{323}, 1324 (2008)
[arXiv:0707.0847 [gr-qc]].
%
\bibitem{Liao:2018sci}
J.~H.~Liao, S.~P.~Miao and R.~P.~Woodard,
{\it Cosmological Coleman-Weinberg Potentials and Inflation},
Phys. Rev. D\textbf{99}, no.10, 103522 (2019)
[arXiv:1806.02533 [gr-qc]].
%  
 \bibitem{Miao:2020zeh}
S.~P.~Miao, L.~Tan and R.~P.~Woodard,
{\it Bose\textendash{}Fermi cancellation of cosmological Coleman\textendash{}Weinberg potentials},
Class. Quant. Grav.\textbf{37}, no.16, 165007 (2020)
[arXiv:2003.03752 [gr-qc]].
%
\bibitem{Glavan:2019uni}
D.~Glavan and G.~Rigopoulos,
{\it One-loop electromagnetic correlators of SQED in power-law inflation},
JCAP\textbf{02}, 021 (2021)
[arXiv:1909.11741 [gr-qc]].
%
\bibitem{Karakaya:2019vwg} 
  G.~Karakaya and V.~K.~Onemli,
{\it Quantum Fluctuations of a Self-interacting Inflaton},
  arXiv:1912.07963.
%
\bibitem{Cabrer:2007xm}
J.~A.~Cabrer and D.~Espriu,
{\it Secular effects on inflation from one-loop quantum gravity},
Phys. Lett. B \textbf{663}, 361-366 (2008)
[arXiv:0710.0855 [gr-qc]].
%
\bibitem{Prokopec:2003qd}
T.~Prokopec and R.~P.~Woodard,
{\it Production of massless fermions during inflation},
JHEP \textbf{10}, 059 (2003)
[arXiv:astro-ph/0309593 [astro-ph]].
%
\bibitem{Boran:2017fsx}
S.~Boran, E.~O.~Kahya and S.~Park,
{\it Quantum gravity corrections to the conformally coupled scalar self-mass-squared on de Sitter background. II. Kinetic conformal cross terms},
Phys. Rev. D \textbf{96}, no.2, 025001 (2017)
[arXiv:1704.05880 [gr-qc]].
%

%%%%%%%%%%%%%%%%%%%%%%%%%%%%%%%%%%%%%%%%%%%%%%%%%%%%%%%%%%%%%%%%%%%%%%%%%%%%%%%%%%%%%%%%%%%%%%%%%%%%%%
%for resummation
%
\bibitem{Moreau:2018ena}
G.~Moreau and J.~Serreau,
{\it Backreaction of superhorizon scalar field fluctuations on a de Sitter geometry: A renormalization group perspective},
Phys.~Rev.~D\textbf{99}, no.2, 025011 (2019)
[arXiv:1809.03969 [hep-th]].
%
\bibitem{Moreau:2018lmz} 
G.~Moreau and J.~Serreau,
{\it Stability of de Sitter spacetime against infrared quantum scalar field fluctuations},
Phys.\ Rev.\ Lett.\  {\bf 122}, no. 1, 011302 (2019)
[arXiv:1808.00338 [hep-th]].
%
\bibitem{Gautier:2015pca}
F.~Gautier and J.~Serreau,
{\it Scalar field correlator in de Sitter space at next-to-leading order in a 1/N expansion},
Phys.~Rev.~D\textbf{92}, no.10, 105035 (2015)
[arXiv:1509.05546 [hep-th]].
%
\bibitem{Serreau:2013eoa}
J.~Serreau,
{\it Renormalization group flow and symmetry restoration in de Sitter space},
Phys.~Lett.~B\textbf{730}, 271 (2014)
[arXiv:1306.3846 [hep-th]].
%
\bibitem{Serreau:2013koa}
J.~Serreau,
{\it Nonperturbative infrared enhancement of nonGaussian correlators in de Sitter space},
Phys.~Lett.~B\textbf{728}, 380 (2014)
[arXiv:1302.6365 [hep-th]].
%
\bibitem{Serreau:2013psa}
J.~Serreau and R.~Parentani,
{\it Nonperturbative resummation of de Sitter infrared logarithms in the large-N limit},
Phys.~Rev.~D\textbf{87}, 085012 (2013)
[arXiv:1302.3262 [hep-th]].
%
\bibitem{Ferreira:2017ogo}
R.~Z.~Ferreira, M.~Sandora and M.~S.~Sloth,
{\it Patient Observers and Non-perturbative Infrared Dynamics in Inflation},
JCAP\textbf{02}, 055 (2018)
[arXiv:1703.10162 [hep-th]].
%

\bibitem{Weinberg}
S. Weinberg, 
{\it Quantum contributions to cosmological correlations}, Phys. Rev. D \textbf{72}, 043514 (2005).


\bibitem{Burgess:2009bs}
C.~P.~Burgess, L.~Leblond, R.~Holman and S.~Shandera,
{\it Super-Hubble de Sitter Fluctuations and the Dynamical RG},
JCAP\textbf{03}, 033 (2010)
[arXiv:0912.1608 [hep-th]].
%
\bibitem{Burgess:2015ajz}
C.~P.~Burgess, R.~Holman and G.~Tasinato,
{\it Open EFTs, IR effects $\&$ late-time resummations: systematic corrections in stochastic inflation},
JHEP\textbf{01}, 153 (2016)
[arXiv:1512.00169 [gr-qc]].
%
\bibitem{Youssef:2013by}
A.~Youssef and D.~Kreimer,
{\it Resummation of infrared logarithms in de Sitter space via Dyson-Schwinger equations: the ladder-rainbow approximation},
Phys. Rev. D\textbf{89}, 124021 (2014)
[arXiv:1301.3205 [gr-qc]].
%
\bibitem{Baumgart:2019clc} 
M.~Baumgart and R.~Sundrum,
{\it De Sitter Diagrammar and the Resummation of Time},
  arXiv:1912.09502.
%
\bibitem{Kitamoto:2018dek}
H.~Kitamoto,
{\it Infrared resummation for derivative interactions in de Sitter space},
Phys.~Rev.~D\textbf{100}, no.2, 025020 (2019)
[arXiv:1811.01830 [hep-th]].
%
\bibitem{Kamenshchik:2020yyn}
A.~Y.~Kamenshchik and T.~Vardanyan,
{\it Renormalization group inspired autonomous equations for secular effects in de Sitter space},
Phys. Rev. D\textbf{102}, no.6, 065010 (2020)
[arXiv:2005.02504 [hep-th]].

\bibitem{Kamenshchik:2021tjh}
A.~Y.~Kamenshchik, A.~A.~Starobinsky and T.~Vardanyan,
{\it Massive scalar field in de Sitter spacetime: a two-loop calculation and a comparison with the stochastic approach},
Eur. Phys. J. C \textbf{82}, no.4, 345 (2022).

\bibitem{Tsamis:2005hd}
N.~C.~Tsamis and R.~P.~Woodard,
{\it Stochastic quantum gravitational inflation},
Nucl. Phys. B \textbf{724}, 295-328 (2005)
[arXiv:gr-qc/0505115 [gr-qc]].
%

\bibitem{Bhattacharya:2022aqi}
S.~Bhattacharya,
{\it Massless minimal quantum scalar field with an asymmetric self interaction in de Sitter spacetime},
JCAP \textbf{09}, 041 (2022)
[arXiv:2202.01593 [hep-th]].
%

\bibitem{Bhattacharya:2022wjl}
S.~Bhattacharya and N.~Joshi,
{\it Non-perturbative analysis for a massless minimal quantum scalar with V(\ensuremath{\phi}) = \ensuremath{\lambda}\ensuremath{\phi} $^{4}$/4! + \ensuremath{\beta}\ensuremath{\phi} $^{3}$/3! in the inflationary de~Sitter spacetime},
JCAP \textbf{03}, 058 (2023)
[arXiv:2211.12027 [hep-th]].


\bibitem{Bhattacharya:2023yhx}
S.~Bhattacharya and M.~D.~Choudhury,
{\it Non-perturbative $\langle \phi \rangle$, $\langle \phi^2 \rangle$ and the dynamically generated scalar mass with Yukawa interaction in the inflationary de Sitter spacetime},
[arXiv:2308.11384 [hep-th]].


\bibitem{Bhattacharya:2023xvd}
S.~Bhattacharya, N.~Joshi and K.~Roy,
{\it Resummation of local and non-local scalar self energies via the Schwinger-Dyson equation in de Sitter spacetime},
[arXiv:2310.19436 [hep-th]].




\bibitem{Glavan:2021adm}
D.~Glavan, S.~P.~Miao, T.~Prokopec and R.~P.~Woodard,
{\it Large logarithms from quantum gravitational corrections to a massless, minimally coupled scalar on de Sitter},
JHEP \textbf{03} (2022), 088
[arXiv:2112.00959 [gr-qc]].


\bibitem{Litos:2023nvj}
C.~Litos, R.~P.~Woodard and B.~Yesilyurt,
{\it Large inflationary logarithms in a nontrivial nonlinear sigma model},
Phys. Rev. D \textbf{108} (2023) no.6, 065001
[arXiv:2306.15486 [gr-qc]].


\bibitem{Glavan:2023lvw}
D.~Glavan and T.~Prokopec,
{\it When tadpoles matter: one-loop corrections for spectator Higgs in inflation},
JHEP \textbf{10} (2023), 063
[arXiv:2306.11162 [hep-ph]].

\bibitem{Glavan:2023tet}
D.~Glavan, S.~P.~Miao, T.~Prokopec and R.~P.~Woodard,
{\it Explaining large electromagnetic logarithms from loops of inflationary gravitons}
JHEP \textbf{08} (2023), 195
[arXiv:2307.09386 [gr-qc]].


\bibitem{Miao:2024nsz}
S.~P.~Miao, N.~C.~Tsamis and R.~P.~Woodard,
{\it Summing Gravitational Effects from Loops of Inflationary Scalars},
[arXiv:2405.01024 [gr-qc]].

  





%
%%%%%%%%%%%%%%%%%%%%%%%%%%%%%%%%%%%%%%%%%%%%%%%%%%%%%%%%%%%%%%%%%%%%%%%%%%%%%%%%%%%%%%%%%%%%%%%%%%%%%%%% 


%
\bibitem{Starobinsky:1986fx}
A.~A.~Starobinsky,
{\it Stochastic de sitter (inflationary) stage in the early universe},
Lect. Notes Phys. \textbf{246}, 107-126 (1986).
%
\bibitem{Starobinsky:1994bd}
A.~A.~Starobinsky and J.~Yokoyama,
{\it Equilibrium state of a selfinteracting scalar field in the De Sitter background},
Phys. Rev. D \textbf{50}, 6357-6368 (1994) [arXiv:astro-ph/9407016 [astro-ph]].
%
\bibitem{Cho:2015pwa}
G.~Cho, C.~H.~Kim and H.~Kitamoto,
{\it Stochastic Dynamics of Infrared Fluctuations in Accelerating Universe},
doi:10.1142/9789813203952\_0018
[arXiv:1508.07877 [hep-th]].
%
\bibitem{Prokopec:2015owa}
T.~Prokopec,
{\it Late time solution for interacting scalar in accelerating spaces},
JCAP \textbf{11}, 016 (2015)
[arXiv:1508.07874 [gr-qc]].
%
\bibitem{Garbrecht:2013coa}
B.~Garbrecht, G.~Rigopoulos and Y.~Zhu,
{\it Infrared correlations in de Sitter space: Field theoretic versus stochastic approach},
Phys. Rev. D \textbf{89}, 063506 (2014)
[arXiv:1310.0367 [hep-th]].
%
\bibitem{Vennin:2015hra}
V.~Vennin and A.~A.~Starobinsky,
{\it Correlation Functions in Stochastic Inflation},
Eur. Phys. J. C \textbf{75}, 413 (2015).
%
\bibitem{Cruces:2022imf}
D.~Cruces,
{\it Review on Stochastic Approach to Inflation},
Universe \textbf{8}, no.6, 334 (2022)
[arXiv:2203.13852 [gr-qc]].
%
\bibitem{Finelli:2008zg}
F.~Finelli, G.~Marozzi, A.~A.~Starobinsky, G.~P.~Vacca and G.~Venturi,
{\it Generation of fluctuations during inflation: Comparison of stochastic and field-theoretic approaches},
Phys.~Rev.~D\textbf{79}, 044007 (2009)
[arXiv:0808.1786 [hep-th]].
%
\bibitem{Markkanen:2019kpv}
T.~Markkanen, A.~Rajantie, S.~Stopyra and T.~Tenkanen,
{\it Scalar correlation functions in de Sitter space from the stochastic spectral expansion},
JCAP\textbf{08}, 001 (2019)
[arXiv:1904.11917 [gr-qc]].
%
\bibitem{Markkanen:2020bfc}
T.~Markkanen and A.~Rajantie,
{\it Scalar correlation functions for a double-well potential in de Sitter space},
JCAP\textbf{03}, 049 (2020)
[arXiv:2001.04494 [gr-qc]].


\bibitem{Enqvist:2017kzh}
K.~Enqvist, R.~J.~Hardwick, T.~Tenkanen, V.~Vennin and D.~Wands,
{\it A novel way to determine the scale of inflation},
JCAP \textbf{02}, 006 (2018) [arXiv:1711.07344 [astro-ph.CO]].







%
\bibitem{davis}
R.~L.~Davis,
{\it On dynamical mass generation in de Sitter space},
Phys. Rev. D \textbf{45}, 2155 (1992).


\bibitem{Beneke:2012kn}
M.~Beneke and P.~Moch,
{\it On dynamical mass generation in Euclidean de Sitter space},
Phys. Rev. D \textbf{87}, 064018 (2013)
[arXiv:1212.3058 [hep-th]].




%%%%%%%%%%%%%%%%%%%%%%%%%%%%%%%%%%%%%%%%%%%%%%%%%%%%%%%%%%%%%%%%%%%%%%%%%%%%%%%%%%%%%%%%%%%%%%%%%%%%%%%
  



%%%%%%%%%%%%%%%%%%%%%%%%%%%%%%%%%%%%%%%%%%%%%%%%%%%%%%%%%%%%%%%%%%%%%%%%%%%%%%%%%%%%%%%%%%%%%%%%%%%%%%%%%in-in formalism


\bibitem{Calzetta}
E. Calzetta and B. L. Hu, 
{\it Closed Time-Path Functional Formalism in Curved Spacetime: Application to
Cosmological Back-Reaction Problems}, Phys. Rev. D \textbf{35}, 495 (1987).

\bibitem{Hu}Adshead
E. Calzetta and B. L. Hu, 
{\it Nonequilibrium quantum fields: Closed-time-path effective action, Wigner function, and Boltzmann equation}, Phys. Rev. D \textbf{37}, 2878 (1988).



\bibitem{Adshead}
P. Adshead, R. Easther and E. A. Lim, {\it The `in-in' Formalism and Cosmological Perturbations}, 
Phys. Rev. D \textbf{80}, 083521 (2009) [arXiv:0904.4207 [hep-th]].


\bibitem{Peskin}
M.~E.~Peskin and D.~V.~Schroeder, {\it An introduction to quantum field theory}, Addison-Wesley, USA (1995). 


\bibitem{Parker:2009uva}
L.~E.~Parker and D.~Toms,
{\it Quantum Field Theory in Curved Spacetime: Quantized Field and Gravity},
Cambridge University Press (2009)
doi:10.1017/CBO9780511813924



%%%%%%%%%%%%%%%%%%%%%%%%%%%%%%%%%%%%%%%%%%
%%%%%%%%%%%%%%%%%%%%%%%%%%%%%%%%%%%%%%%%

\end{thebibliography}
\end{document}